\documentclass[11pt,a4paper]{article} 
\pdfoutput=1
\usepackage[no-natbib-sort]{my-jheppub}

\usepackage{amsmath, amssymb}
\usepackage{mathpazo}
\usepackage{environ}
\usepackage{mathrsfs}
\usepackage{array,arydshln}

\usepackage{graphicx,epsfig}
\usepackage{epic}
\usepackage{youngtab}
\usepackage{float}
\usepackage{color}
\definecolor{maroon}{rgb}{0.8,0.3,0.}

\usepackage{slashed}
\usepackage[nodayofweek]{date time}

\usepackage{hyperref}

\usepackage{aurical}
\usepackage[T1]{fontenc}

\newcommand{\be}{\begin{equation}}
\newcommand{\ee}{\end{equation}}

\newcommand{\ads}{AdS$_5\times S^5$\ }

\newcommand{\mc}{\mathcal }

\newcommand{\N}{\mathcal{N}}

\def \TT  {{\rm T}}

\newcommand{\mb}{\mathbb }

\newcommand{\bv}{\begin{verbatim}}
\newcommand{\ev}{\end{verbatim}}

\def \del{ \partial}
\def \la {\label}
\newcommand{\rf}[1]{(\ref{#1})}
\def\ov{\over}
\def\no{\nonumber} \def \aa {{\rm a}}
\def \ci {\cite}

\def \m {\mu}\def \n {\nu} 
\def \ed {\end{document}}
\def \l {\lambda} \def \r {\rho} 

\def \foot {\footnote}
\def \b {\beta} 
\def \dd {{\rm d}} 
 
\def \tr {{\rm tr}}
 
\def \vp {\varphi} 
 \def \ha {{{1 \ov 2}}}

\def \cc {{\rm c}}
\def \aa {{\rm a} }

   \def \tr  {{\rm tr }}   
 
\def \ads {AdS$_{5}$\ }
\def \te {\textstyle} \def \iffa {\iffalse} 
\def \ha {{\te {1 \ov 2}}}
\def  \ba { \begin{align} }
\def  \ea { \end{align} }

\def \cc    {{\rm c}} 
\def \aa  {{\rm a}}
\def \kk  {{\rm k}}
\def \ep {\epsilon}
\def \k {\kappa} \def \r {\rho}

\def \edd {\end{document}} 
\def \td {\tilde}

\def \SS {{\rm V}}

\def \iffa  {\iffalse}
\title{Conformal anomaly c-coefficients  of    superconformal  6d theories }
\author[a]{Matteo Beccaria\footnote{matteo.beccaria@le.infn.it}} 
\author[b]{ and \ \ Arkady A. Tseytlin\footnote{Also at Lebedev Institute, Moscow. \ \ tseytlin@imperial.ac.uk}} 

\abstract{ 
We propose  general relations between the conformal anomaly and the chiral (R-symmetry and gravitational)
anomaly coefficients in 6d $(1,0)$ superconformal theories. 
The suggested   expressions   for the  three  type  B  conformal anomaly $\cc_i$-coefficients  
  complement  the  expression  for the type A anomaly a-coefficient found in   
   \href{http://arxiv.org/abs/1506.03807}{arXiv:1506.03807}.
We check  them   on   several examples --
  the standard  $(1,0)$  hyper  and tensor multiplets as well as  some
  higher derivative  short  multiplets containing vector fields  that generalize the 
  superconformal  6d  vector multiplet discussed in 
   \href{http://arxiv.org/abs/1506.08727}{arXiv:1506.08727}.
We  also consider  a   family of   higher derivative    superconformal   $(2,0)$   6d multiplets 
  associated  to  7d  multiplets  
in the KK spectrum of 11d supergravity compactified on $S^{4}$. 
 In particular, we    prove that  (2,0) 6d conformal supergravity
  coupled to  26    tensor multiplets  is  free of all chiral and  conformal anomalies. 
  We    discuss   some  interacting  $(1,0)$ superconformal theories, predicting the $\cc_i$-coefficients
  for the   "E-string"  theory 
 on multiple    M5-branes  at  $E_8$  9-brane    and  for the 
   theory   describing    M5-branes at  an  orbifold singularity  $\mathbb{C}^{2}/\Gamma$. 
 Finally, we  elaborate on     holographic computation  of subleading corrections  to   conformal anomaly coefficients coming from 
 $R^2+R^3$   terms in 7d effective action,  revisiting, in particular,  the (2,0) theory case. 
\vfill }

\affiliation[a]{Dipartimento di Matematica e Fisica Ennio De Giorgi,\\
Universit\`a del Salento \& INFN, Via Arnesano, 73100 Lecce, 
Italy} 

\affiliation[b]{The Blackett Laboratory, Imperial College, London SW7 2AZ, U.K.}

\allowdisplaybreaks

\begin{document}

 \begin{flushright}\small{Imperial-TP-AT-2015-{06}}\end{flushright}				

\maketitle
\flushbottom

\def \kk {{\rm k}}
\def \qq {{\rm q}}
\newcommand{\hh}{{\textstyle\frac{1}{2}}}
\def \ep {\epsilon } 
\def \adss {AdS$_5 \times S^5$ }
\def \adsss {AdS$_7 \times S^4$ }
\def \adse {AdS$_7$ }
\def \ads {AdS$_5$ }
\def \adt {AdS$_3$ }
\def \bh {{\bar h}} \def \bfh {{\bf{h}}}
\def \G {\Gamma} 
\def \ed {\end{document}}
\def \a{\alpha} \def \b {\beta} \def \na {\nabla} \def \g{\gamma} \def \d {\delta} 
\def \W {{\cal W}} \def \const {{\rm const}}

\def \rp {{\rm p}}

\def \CSG {{\rm CSG}}

\def \csg {CSG$_p$ \ }

\section{Introduction and summary}

The aim of this paper is to  find general  relations between the conformal anomaly  and chiral (R-symmetry and gravitational)  anomaly 
coefficients   in 6d  $(1,0)$ superconformal theories. 
Recently,  the  expression for  the   conformal anomaly   a-coefficient   was   suggested in   \cite{Cordova:2015fha}. 
Here we shall  do the same  for the  conformal anomaly  c-coefficients. 

In 4d  $\N=1$ superconformal theories  there  are   linear relations between  the $\aa$ and $\cc$  coefficients   in the conformal anomaly \ci{Duff:1993wm}\foot{Here  and below $C$ is the Weyl tensor, $E_d$ is the Euler density 
 and we   omit  total  derivative terms in $\langle T\rangle$.} 
\be \la{1}
A_4 \equiv  (4 \pi)^2 \langle T\rangle  = - \aa\, E_4   + \cc\, C^2 
\ee
and the coefficients $\a,\b$  in the $U(1)$  R-symmetry and mixed  gravitational  anomaly 6-form
polynomial  \ci{Anselmi:1997ys,Aharony:2007dj}
(here   $R^2 \equiv R\wedge R $, {\em etc.})\foot{The manifestly supersymmetric  description of  $\N=1$ superconformal  anomalies 
was given in   \cite{Buchbinder:1986im}.}
\ba \la{2} 
&\te \aa= \frac{3}{32}(3\,\a-\b)  \ , \qquad \qquad 
\cc= \frac{1}{32}(9\,\a-5\,\b) \ , \\
 \la{3}
&{\cal I}_6= \te { 1 \ov 3!}  \big( \a\,   c_1^3 + \b\,  c_1 \rp_1\big) \ , \ \ \qquad  \ \ \ \ 
c_1 = \tr\,  F\ , \ \ \ \ \    \rp_1 = -\frac{1}{2}\,\tr\,  R^2 \ . 
\end{align}
\iffa
\be
\label{A.1}
\aa  \sim 3\,\a-\b,\qquad 
\cc  \sim 9\,\a-5\,\b, \qquad 
\b = \text{Tr}U(1)_{R}, \ \ \ \a = \text{Tr}
U(1)^{3}_{R}.
\ee
\fi 
The conformal anomaly of a classically  Weyl invariant theory in 6d has the following 
general 
 form  \cite{Bonora:1985cq,Deser:1993yx,Bastianelli:2000hi}
%
\be
\label{4}
 A_6 \equiv (4\pi)^{3} \langle T\rangle = 
  -\aa\,E_{6}+ W_6 \ , \ \ \ \ \ \  \ \ \     W_6=\cc_{1}\,I_{1}+\cc_{2}\,I_{2}+\cc_{3}\,I_{3}\   . 
\ee
Here  
$E_{6} = \ep_6 \ep_6 RRR$ is the Euler density in six dimensions
   and 
$ W_6$   is a combination of  the  three independent   Weyl invariants 
$I_1  = C_{amnb}C^{mpqn}C_{pabq}$, \ \  $I_2 = C_{abmn}C^{mnpq}C_{pqab},$\ \ $ I_3 = C_{mnpq} \na^2 C^{mnpq} + ... $ 
(for details see  \cite{Bastianelli:2000hi}).\foot{The   sign of  $E_6$  to be used here (we shall assume Euclidean signature) 
is opposite to the  one in 
 \cite{Bastianelli:2000hi,Beccaria:2015uta}  but the sign of the $\aa$-coefficient will    be the same, i.e.    negative 
 for  standard  unitary scalar,  spinor or  tensor field.  We  choose not to reverse 
 the sign of $\aa$  compared to the 4d  case \rf{1}, {\em i.e.} not to include   an  extra $(-1)^{d/2}$   factor 
    (as is done, {\em e.g.},   in \cite{Fei:2015oha}   in discussion of free energy on a sphere)
 so that  the  $\aa$-coefficient   is  always  directly proportional to the  coefficient 
 of the  logarithmic UV divergence (see, {\em e.g.}, \cite{Tseytlin:2013fca}). }
Thus, in general, in 6d there are 4 independent conformal anomaly coefficients.\foot{The Weyl anomaly may be related to the correlation functions of 
the stress tensor on a flat background \cite{Deser:1999zv}. 
In 4d the  $\aa$-coefficient   is related to the 3-point  function $\langle TTT\rangle$ and 
$\cc$   is proportional to the coefficient  in the  2-point function $\langle TT\rangle$.
In 6d, the   $\aa$-coefficient 
 is related to the 4-point function $\langle TTTT\rangle$. The  coefficient 
 $\cc_{3}$ is  related 
to  $\langle TT\rangle$, while $\cc_{1}$ and $\cc_{2}$ are related to the two free parameters
in  $\langle TTT\rangle$ (in 6d this   3-point function has three 
parameters, but one of them is related to the 2-point function by a conformal 
Ward identity \cite{Osborn:1993cr,Erdmenger:1996yc,Bastianelli:1999ab}).} 
Let us note that  on a Ricci flat background  one has the identities
\be\la{5}
E_{6} = 32\,(2\,I_{1}+I_{2})\ , \qquad\qquad  I_{3} = 4\,I_{1}-I_{2}\ ,
\ee
so that \rf{4} takes  the form 
\be
\label{6}
 A_{6}\big|_{R_{mn}=0}  =\te -\big[\aa - \frac{1}{192} (\cc_1 + 4 \cc_{2})\big]\, E_{6}
 +(\cc_{1}-2\cc_{2}+6\cc_{3})\,
I_{1}\ .
\ee
In the presence of   $(1,0)$  supersymmetry  one expects that the three  Weyl invariants $I_i$   are 
bosonic parts of only two possible 6d superinvariants, {\em i.e.}  the coefficients  $\cc_i$   should  satisfy  one linear relation. 
As  suggested by   free-theory  examples ($(1,0)$ scalar and tensor  multiplet)   \cite{Bastianelli:2000hi} 
and holography based  strong-coupling arguments  \ci{Kulaxizi:2009pz},   this  relation   should be\foot{This relation appeared also in 
other contexts  in 
\cite{Hofman:2008ar,Safdi:2012sn,Bueno:2015lza}.}  
\be \la{7}
(1,0): \qquad 
 q_1\equiv \cc_1 -  2 \cc_2 + 6 \cc_3 =0 
\ .  
\ee
Then, \rf{6} implies  that  in this case the  conformal anomaly   on a Ricci flat background   is 
simply proportional   to the Euler density  (and thus  the  integrated or 
scale anomaly vanishes  on an asymptotically flat space). 

In the  case of  $(2,0)$    supersymmetry, the invariants  $I_i$  should be   part of a single superinvariant ${\cal W}_6$ 
  so that $\cc_i$ 
should be subject  
 to 
 one additional   constraint 
 \be  \la{8} 
(2,0): \qquad  q_2 \equiv  \cc_1 - 4 \cc_2=0 \ , \ \ \ \ \ \ \   {\rm i.e.}\quad   \ \ \ \ \    \cc_1 =4 \cc_2 = -12 \cc_3 \equiv 96 \cc \ .  \ee 
 In this case   there is  only one independent $\cc$-coefficient, {\em i.e.} 
 \be
\label{9}
 (2,0): \qquad  A_6 = 
  -\aa\,E_{6}+\cc\,   \W_6 \ , \ \ \ \ \ \  \ \ \     \W_6\equiv 96 \,I_{1}+24 \,I_{2} -8\,I_{3}\   . 
\ee
This   expectation is supported   by the  results  for the  anomaly of a free $(2,0)$   tensor multiplet  in \cite{Bastianelli:2000hi}   and  for 
the strong coupling (large $N$) limit   of interacting  $(2,0)$  theory  \ci{Henningson:1998gx}. 
Then on Ricci flat  background one gets from \rf{6} 
\be \la{10} 
(2,0): \qquad A_{6}\big|_{R_{mn}=0}  = \,\big(\cc-\aa\big)\, E_{6} \ , 
\ee
which is the analog of  the  familiar 
$A_{4}\big|_{R_{mn}=0}  = (\cc-\aa)\,  E_{4}$ relation  in 4d  following from \rf{1}. 

Since  like  4d  anomalies   the 6d anomalies form a supersymmetry multiplet \ci{Howe:1983fr,Manvelyan:2000ef,Manvelyan:2003gc}
  one expects  to find  linear relations between their coefficients   analogous to   \rf{2}. 
   The  6d  chiral ($SU(2)$ R-symmetry and gravitational) 
  anomaly 8-form polynomial 
   has the following general structure \ci{Frampton:1983ah,AlvarezGaume:1983ig,Zumino:1983rz,Faddeev:1985iz} 
   \ba 
   \la{11}
 &  \mc I_{8} =\te  \frac{1}{4!}\big(\alpha\,c_{2}^{2}+\beta\,c_{2}\,\rp_{1}+
\gamma\,\rp_{1}^{2} +\delta\,\rp_{2} \big) \ ,\\
  &  c_2 = \tr\,  F^2 \ , \ \ \ \qquad \ \ \    \te \rp_2 = -\frac{1}{4}\,\tr\,   R^4 +\frac{1}{8}\,(\tr\,R^{2})^{2}\ , \la{111}
\end{align}
  where $c_1$ and $\rp_1$ are defined as in \rf{3} and  $\vec \a = (\a,\b, \g, \d)$  
   are numerical coefficients.\foot{The R-symmetry  gauge  bundle  forms 
   $c_1$ and $c_2$   should  not be confused with the   conformal  anomaly 
    coefficients $\cc_1,\cc_2$.} 
    
In    \cite{Cordova:2015fha}  an expression for the  $\aa$-coefficient in \rf{4}  in  $(1,0)$ superconformal 
theory written  in terms of  the  chiral   anomaly coefficients 
$\vec \a$   was found by  determining the 4  parameters  $\vec k$  in the expected linear relation $\aa= \vec k \cdot \vec \a$  
from  several  explicit  examples:\foot{This relation 
 (and similar relations below)  between chiral and conformal anomalies applies 
 for $(1,0)$   supermultiplets;  for $(0,1)$  ones one  should flip the overall sign 
 as  conformal anomaly  is not sensitive to chirality (of fermions  and antisymmetric tensors)   while chiral anomalies are.}
\be
\label{12}
\aa = \te - {1 \ov 72} \big( \alpha-\beta+\gamma 
+\frac{3}{8}\,\delta\big)\ . \ee
Equivalently, this may be written as 
\be \la{13}
\aa= \te  \frac{16}{7} \big( \alpha-\beta+\gamma 
+\frac{3}{8}\,\delta\big)\,  \aa_{\rm T}    \ , \ \ \ \qquad \ \ \ \ \   \ \ \   \aa_{\rm T}  \equiv \aa(T^{(2,0)})= - { 7 \ov 1152} \ , 
\ee
where $ \aa_{\rm T} $ is the value of $\aa$   for the $(2,0)$  tensor multiplet     \cite{Bastianelli:2000hi}
for which $\vec \a= (1,{1\ov 2}, {1\ov 8}, - { 1\ov 2})$. 
  
Here we shall  follow the  same  strategy to  find the analogs of \rf{12}   for the 
 two  independent  conformal  anomaly $\cc_i$-coefficients, {\em e.g.}, 
$\cc_1$ and $\cc_2$  (with $\cc_3$ then given by \rf{7}). Using as input  some  known examples we 
suggest the following expressions
\foot{\la{new-foot}
The general expressions for $c_1$ and $c_2$    that  we  obtained  after using currently  available data 
depend  on one  free  1-parameter  $\xi$:
\be
\notag
\te 
\cc_{1} = -\frac{4}{3}\alpha+ (\frac{6}{7}\,\xi+\frac{146}{63})\beta+
(\frac{4}{7}\xi-\frac{80}{63})\gamma+\xi\delta, \qquad 
\cc_{2} = -\frac{1}{3}\alpha+(
-\frac{3}{14}\xi+\frac{11}{126})\beta+
(
-\frac{1}{7}\xi-\frac{22}{63}
)\gamma + (
-\frac{1}{4}\xi-\frac{1}{2}
)\delta. 
\ee
The specific  form 
in \rf{14} is obtained for the  particular  choice  of $\xi = -\frac{31}{27}$  that will be   
conjectured  to be the   correct one 
 in   section  5   as it  leads to the  relations 
\rf{28} and \rf{30}.
Let us  note that   the  expressions for $\cc_i$   found in recent paper \ci{Yankielowicz:2017xkf}
also belong to the above 1-parameter family but for  a different value of $\xi=  -\frac{8}{9}$.
It remains to  be seen  if the resulting  expressions in   \rf{14}  may be 
used to interpret (at fixed points) 
the RG flow $C$-functions discussed in \cite{Heckman:2015axa}  in terms of the conformal anomaly coefficients.
}
\be
\label{14}
\begin{split}
&\te  \cc_{1} = -\frac{4}{3}\,(\alpha-\beta)-\frac{52}{27}\,\gamma-\frac{31}{27}\,\delta, \qquad \qquad 
\cc_{2} = -\frac{1}{3}\,(\alpha-\beta)-\frac{5}{27}\,\gamma-\frac{23}{108}\,\delta \ ,\\
& \qquad  \ \ \ \ \ \ \  \qquad  \qquad \qquad  \te  \cc_3 =- {1\ov 6} ( \cc_1 - 2 \cc_2) \ . 
\end{split}
\ee
Like $\aa$ in \rf{12}  all $\cc_i$ then  depend  on $\a,\b$ only  through  
 $\a-\b$, suggesting that  $c^2_2 - c_2 \rp_1$  combination in the anomaly \rf{11} 
 is  part of  $(1,0)$ superinvariant  related to conformal anomaly. 
 Also, we    observe that  
 \be \la{15}\te   \cc_{1}-4\cc_{2} = -\frac{8}{27}(4\gamma+\delta) \ , \ee
 so that in the special $(2,0)$  supersymmetry case when \rf{8} is expected to be satisfied 
   we should have 
 \be \la{16}
(2,0):\qquad   \qquad \te  \gamma = -\tfrac{1}{4}\,\delta \ . \ee
This is consistent with the fact that $\rp_2 - {1\ov 4} \rp_1^2$ in \rf{11}   should 
form  part of   associated   superinvariant.\foot{This 8-form multiplied by $C_3$-potential   is  bosonic part of a 11d  $\ep_{11} C_3 RRRR$   superinvariant.
}
  Also, in the $(2,0)$ case the   coefficients  $\b$ and $\g$  in \rf{11} 
  appear to be related by
  \be \la{166}
(2,0):\qquad   \qquad \te  \b = 4 \gamma  \ . \ee
Then it follows that  in the (2,0) case  (see \rf{8},\rf{12},\rf{14},\rf{16})  
 \ba \la{17}
 (2,0):  \qquad \qquad 
 &\aa= \te - {1 \ov 72} \big( \alpha+\frac{9}{8}\,\delta\big)\ ,   \qquad 
 \cc= \te - {1 \ov 72} \big( \alpha  + \frac{3}{2}\,\delta\big)\ ,\\
& \qquad   \qquad   \cc_1 = 4 \cc_2 = -12 \cc_3 = 96 \cc \ , \la{777} \\
 &\qquad   \qquad \qquad  \ \ \  \
\te  \aa-\cc = {1 \ov 192} \delta \ . \la{77}
 \end{align}
The    relation \rf{77} allows one to determine $\cc$ from the knowledge of $\aa$-coefficient and the gravitational anomaly  
coefficient $\delta$.

Below, we shall be  discussing 1-loop conformal  anomalies in  higher derivative theories.\foot{In general, 
the properties    of  anomalies  in higher-derivative theories  (and comparison   with their derivation in the standard 
 unitary theories)  deserves  a thorough investigation that we postpone till future work.}
In general, the conformal anomaly in the trace  of stress tensor on a curved background is  related to 
a logarithmic UV  divergence  and thus  it is determined  by the corresponding heat kernel (Seeley)   coefficient. 
This   relation is ``kinematical" and is expected to apply  regardless the  order of the  differential operator  involved 
or metric signature  (and thus  also other  issues  like unitarity).
One important assumption will  be the existence of Weyl-covariant   generalisations of   flat-space conformal  higher derivative   operators 
that  have the correct gauge invariance at least on  conformally flat and   Ricci-flat 
backgrounds  (cf. \cite{Tseytlin:2013jya}).\foot{We will  considering 
only universal, {\em i.e.}   regularisation (or scheme)  independent terms in 
6d conformal  anomaly  that cannot be changed by adding local counterterms.}
As for chiral anomalies, they are essentially the same (modulo ghost  counting, {\em etc.}) 
 for  the   standard and  higher-derivative operators 
(see, {\em e.g.}, \ci{AlvarezGaume:1983ig,Romer:1985yg,Smilga:2006ax,Carrasco:2013ypa}).

\

The structure of  the rest of this  paper is as follows. 
In section 2   we shall    summarize the results for conformal and chiral anomalies 
of the standard (1,0) hyper $S^{(1,0)}$ and tensor  $T^{(1,0)} $  multiplets   and also  for the  6d 
 conformal  analog $V^{(1,0)}$
   of the Maxwell  multiplet   which contains  4-derivative vector,  3-derivative   spinor   and 2-derivative scalar  fields.
    This  will provide evidence for the relation \rf{14}. 

 In section 3   we shall consider a family  $V_p^{(1,0)}$ of   higher  derivative 
 ``massive"     (1,0)    conformal   multiplets generalizing the ``massless"  (gauge-invariant) 
 vector multiplet 
 $V^{(1,0)}$.  They appear in tensor product  of $p$ copies of $S^{(1,0)}$   scalar multiplets. 
     We shall  independently compute their chiral  and conformal 
 anomalies  and  provide  an additional check of \rf{14}. 
 
 In section 4 we shall  compute the anomalies of a  family \csg  of (2,0)  6d  multiplets that 
  generalize the (2,0) conformal supergravity multiplet (corresponding to $p=2$ case). These   are associated 
    to 7d multiplets that appear in the KK spectrum of 11d supergravity  compactified on $S^4$, i.e. appear in tensor product  of $p$ copies of $T^{(2,0)}$  tensor multiplets. 
 We verify the relation \rf{13}  and apply \rf{17} to compute the corresponding $\cc$-anomaly.
 We show, in  particular, that the system of (2,0)  conformal supergravity coupled  to 26 tensor multiplets is  chiral and conformal 
  anomaly free. 
 
 In section 5    we  shall apply the relations \rf{14} to compute $\cc_i$-coefficients of  some  interacting 
 (1,0) superconformal 6d theories for which the chiral anomaly coefficients are known:  the 
 E-string  theory 
 on multiple    M5-branes  at  $E_8$  9-brane    and   the 
   theory  on   M5-branes at  an  orbifold singularity  $\mathbb{C}^{2}/\Gamma$. 
   We observe some relations between  the $\cc_i$-coefficients  that  indirectly support the consistency of our suggested 
   expressions \rf{14}.
 
  Finally,  in section 6   we  shall   consider   
  the  AdS/CFT  based  computation  of subleading corrections  to  the 6d  conformal anomaly coefficients coming from 
 $R^2+R^3$   terms in 7d effective action   discussing, in particular,  the   supersymmetry constraints \rf{7} and \rf{8}
 and $1/N^2$   corrections to conformal anomaly in (2,0) theory.

In Appendix \ref{A}
we shall   compute the 
conformal anomaly   coefficients   for the   higher-derivative vector multiplet 
$V^{(1,0)}$  on a  Ricci flat background. 
Appendix \ref{C}   will be devoted to the  computation of  chiral anomaly  coefficients 
for the  \csg    multiplets. We shall also  give a separate  discussion 
 of   the   anomalies of the  (2,0)   and (1,0)   conformal supergravities. 
In Appendix \ref{D}  we will  propose an expression for  the  $S^5$  Casimir energy   $E_c$ 
for  (1,0)   multiplets    given   like  \rf{12},\rf{14}   by a linear combination of the chiral anomaly  coefficients.

\iffa

To support \rf{14} we shall first discuss available perturbative  data and then consider 
information available at strong coupling via holography. 
we shall first discuss available 
Discuss  conformal   anomaly of $(2,0)$ conformal supergravity.

give review of count for vector mult   eliminating confusion -- explaining 
yes, indeed in his flow what is relevant is the variation of alpha-beta+gamma, not alpha-beta alone.
Unfortunately there is nothing like dilation actions useful to extract c_i because in the argument exploiting dilaton
the Weyl invariant terms c_i I_i have trivial variation under Weyl transformations, while E_6 has not.
\fi 

\section{Simplest  free $(1,0)$  6d  supermultiplets}
The  conformal anomalies  of  free  2-derivative  scalar $\vp$, Majorana-Weyl (MW)  spinor $\psi$  and (anti)selfdual 
 rank 2 tensor  $T$ were computed in \ci{Bastianelli:2000hi}.
 Combining these fields into $(1,0)$  scalar and tensor   and $(2,0)$  tensor   supermultiplets
 (we indicate physical fields only and  their   chirality) 
 \ba
S^{(1,0)} = 4\varphi+2\psi^{-}, \qquad\qquad  T^{(1,0)} = \varphi+2\psi^{-}+T^{-}\ ,\no  \\ \ \ \ \ \ \ \ \ 
T^{(2,0)} = S^{(1,0)}+T^{(1,0)} = 5\varphi+4\psi^{-}+T^{-} \ , \la{18}
\end{align}
   we  get 
 the corresponding  values of  $\aa$ and $\cc_i$  in \rf{4}. 
 The results are   summarized in Table \ref{T1}. 
\begin{table}[H]
\be
\def\arraystretch{1.3}
\begin{array}{|c|cccc|}
\hline
\mbox{ } & \aa & \cc_{1} & \cc_{2} & \cc_{3} \\
\hline
\varphi &  -\frac{1}{72576} & -\frac{1}{540} & \frac{1}{3024} & \frac{1}{2520} \\
\psi^- & -\frac{191}{1451520} & -\frac{2}{135} & -\frac{1}{630} & \frac{1}{504}\\
T^- & -\frac{221}{40320} & -\frac{143}{540} & -\frac{1189}{15120} &  \frac{1}{56}\\
\hline
S^{(1,0)} & -\frac{11}{34560} & -\frac{1}{27} & -\frac{1}{540} & \frac{1}{180}\\
T^{(1,0)} & -\frac{199}{34560} & -\frac{8}{27} & -\frac{11}{135} & \frac{1}{45}\\
\hline
T^{(2,0)} & -\frac{7}{1152} & -\frac{1}{3} & -\frac{1}{12} & \frac{1}{36}\\
\hline
\end{array}\notag
\ee
\caption{Conformal anomaly  coefficients for free  fields and supermultiplets}
\label{T1}
\end{table}
We observe that for $S^{(1,0)}$ and $T^{(1,0)}$ the relation \rf{7}  is  satisfied, while 
for $T^{(2,0)}$  we have also the relation  \rf{8}. 
It is reasonable to assume that \rf{7}  follows from the requirement of $(1,0)$ supersymmetry 
on the Weyl super-cocycle  and should thus be true 
 in all $(1,0)$ cases, free or interacting. This is also consistent also with what is  implied  by 
 holographic computation of conformal anomaly  \ci{Kulaxizi:2009pz,deBoer:2009pn}. 
 
Let us   also  consider  the standard   6d  $(1,0)$ supersymmetric 
vector multiplet $V_{\rm s}^{(1,0)}$  (which includes    the  2-derivative Maxwell field $V$
and thus is not superconformal)  and  also the higher-derivative superconformal 6d  $(1,0)$  
vector multiplet  $V^{(1,0)}$ that 
includes the 4-derivative vector  $V^{(4)}  $ and 3-derivative spinor   $\psi^{(3)}  $  \cite{Ivanov:2005kz,Beccaria:2015uta}
\be\la{19}
V_{\rm s}^{(1,0)} = 2\psi^{+}+V, \qquad\qquad  V^{(1,0)} = 3\varphi+2\psi^{(3), +}+V^{(4)} \ . 
\ee
The supersymmetric  vector multiplet $V_{\rm s}^{(1,0)}$   may be appearing in low-energy limit 
of a spontaneously broken 6d superconformal theory ({\em e.g.},  from $\phi F^2_{\mu\nu} + ... $   with $
\langle  \phi \rangle  =\const$).  While  the   conformal  anomaly coefficients are not defined for  non-conformal 
$V_{\rm s}^{(1,0)}$   multiplet,    for $V^{(1,0)} $    one finds 
\be 
V^{(1,0)} : \qquad  \te  \aa= {251\ov 34560} 
\ , \ \ \ \ \ \ \qquad 
\cc_1  = {19\ov 27} \ , \ \ \ \  \cc_2 = {91\ov 540} \ , \ \ \ \    \cc_3 = - {11\ov 180} \ .\la{20}\ee
Here $\aa$ can be determined  \ci{Beccaria:2015uta}  from the value of the conformal 
  anomaly on $S^6$ --   either directly in 6d or using the   AdS$_7$  based method 
  \ci{Giombi:2013yva, Giombi:2014iua, Beccaria:2014xda}. The values of $\cc_i$   can be
  computed  starting with  the curved space kinetic operators 
  given in \ci{Beccaria:2015uta}  and     are again consistent with the 
  constraint \rf{7}  (see Appendix \ref{A}).
  
  Let us   now consider the corresponding   chiral anomaly coefficients in \rf{11}.  The only fields that 
  contribute to the chiral anomalies are the Weyl   fermions  and self-dual tensors
  (their values   can be found in  \ci{AlvarezGaume:1983ig}). Also, the  chiral   anomaly contribution 
  of  3-derivative   fermion $\psi^{(3)}  $  is the same   as of the  standard 1-derivative  fermion  $\psi$. 
  This leads  to the  coefficients  summarized   in Table \ref{T2}
  where   we also give the results  for the  $\aa$ coefficient  verifying the relation \rf{13}  
   of \ci{Cordova:2015fha}.
\begin{table}[H]
\be
\def\arraystretch{1.3}
\begin{array}{|c|c|c|c|c|c|}
\hline
 & \alpha & \beta & \gamma & \delta & \aa/\aa_{\rm T}
  \\
 \hline
 S^{(1,0)} & 0 & 0 & \frac{7}{240} & -\frac{1}{60} &  \frac{11}{210} \\
 \hline
 T^{(1,0)} & 1 & \frac{1}{2} & \frac{23}{240} & -\frac{29}{60} & \frac{199}{210}\\
 \hline
 T^{(2,0)} & 1 &  \frac{1}{2}  &  \frac{1}{8}   & -\frac{1}{2}  &  1\\
 \hline
 V^{(1,0)}_{\rm s} & -1 & -\frac{1}{2} & -\frac{7}{240} & \frac{1}{60} &  - 
 \\
 \hline
 V^{(1,0)} & -1 &  -\frac{1}{2}  &  -\frac{7}{240}   & \frac{1}{60}  &  -\frac{251}{210}\\
 \hline
\end{array}\notag
\ee
\caption{Chiral anomaly  coefficients  and $\aa$ coefficient (in units of   $T^{(2,0)}$  value)} 
\label{T2}
\end{table}
Few explanations are in order. The values for   $T^{(2,0)}$   in \rf{18}  are the
sums of the values  for  $S^{(1,0)}$  and $T^{(1,0)}$.
 The chiral  anomalies  of 
$V^{(1,0)}_{\rm s} $  and $V^{(1,0)} $  are the same as they come only from the 
fermions  and are not sensitive to extra $\del^2$ in the kinetic term of $\psi^{(3)}$ in \rf{19}.
The values of $\a,\b$   for $S^{(1,0)}$ are zero  since the corresponding fermions in \rf{18} 
 are singlets of $SU(2)_R$. 
The fermions of $T^{(1,0)}$  form a doublet  of  $SU(2)_R$  as are the fermions of 
$V^{(1,0)}_{\rm s} $  and $V^{(1,0)} $  but their  $\a,\b$ coefficients  differ  in sign due to different chirality assignments in \rf{18},\rf{19} (the antisymmetric tensor of  $T^{(1,0)}$   does not contribute to
$\a,\b$  as it is  singlet of R-symmetry). 
The gravitational anomalies $\gamma, \delta$ 
 of $S^{(1,0)}$  and of $V^{(1,0)}_{\rm s} $ or  $V^{(1,0)} $
are the same  up to sign as they come   from  fermions  of  opposite chirality
(for $T^{(1,0)}$ there are also  additional contributions of antisymmetric tensor).

It  was  noted  in \ci{Cordova:2015fha} that  the relation \rf{13}  formally applied 
to the standard  (scale  invariant 
 but  not  conformally  invariant   vector  multiplet   $V^{(1,0)}_{\rm s} $  for which $\aa$
 is   not defined)  
  leads to 
the  value $\aa= - \frac{251}{210}\aa_{\rm T}$   which has the opposite sign 
to  $\aa$   of $S^{(1,0)}$  and $T^{(1,0)}$.
 As was observed   in \ci{Beccaria:2015uta}, 
this value   corresponds,  in fact,  to  the  higher-derivative superconformal multiplet $V^{(1,0)} $.
It  now appears that the  reason  for this curious  observation   is  simply of  technical  nature 
--  it   follows from the   fact that the chiral anomalies  of   $V^{(1,0)}_{\rm s} $  and $V^{(1,0)} $   are the 
same, and  $V^{(1,0)} $    being higher-derivative (non-unitary)
   happens to have an  opposite  sign of $\aa$-coefficient   compared to   the one of  the 
    standard  unitary   scalar and tensor multiplets. 

Turning now to the   conformal anomaly $\cc_i$-coefficients  we observe that  their values 
in Table \ref{T1}  and   \rf{20}   are indeed consistent  with the relations   \rf{14},\rf{7} 
 where $\a,\b,\g,\d$ are   given  by Table \ref{T2}. 
 The   values  for the three   multiplets $S^{(1,0)}$, $T^{(1,0)}$  and $V^{(1,0)}$  
 are sufficient   to fix  3 out of 4  a priori  unknown  coefficients  in the relations \rf{14} 
 between $\cc_i$ and chiral anomaly coefficients. 
 


\section{Higher derivative  spin 1   superconformal multiplets}

To provide further examples   of anomaly  computation for $(1,0)$   superconformal multiplets  and to support 
 the relations \rf{14}   let us now   consider  a family  higher derivative  analogs 
of the 
 vector multiplet $V^{(1,0)}$. These multiplets,
that we   will denote  as $\SS^{(1,0)}_p$, $p\ge 2$,    will contain  scalars, spinors  and   vectors
with higher-derivative   kinetic terms.  The 
$p=2$ case will  correspond to $V^{(1,0)}$  discussed  above. 

From the point of view
of $OSp(2,6|2)$ representations \cite{Minwalla:1997ka,Dobrev:2002dt,Bhattacharya:2008zy} the hypermultiplet $S^{(1,0)}$
is a doubleton ultra-short representation \cite{Ferrara:2000xg}.\footnote{In general, 
 the relevant superconformal algebras
  are $OSp(8^*\,|\,\mathcal{N})$ with compact R-symmetry $USp(\mathcal{N})$.
Our notation $OSp(2,6|2)$
 implicitly exploits the isomorphism $SO^*(8) = SO(2,6)$ \cite{VanProeyen:1999ni}.
} 
Additional (possibly massive) conformal representations are 
obtained from the tensor product of $p$ copies of $S^{(1,0)}$.
The resulting multiplets $\SS^{(1,0)}_{p}$
are still short but have maximal
spin equal to 1   instead of  $\ha $ as in the $S^{(1,0)}$ case. 
The  structure of these multiplets was  worked out
in  \cite{Gimon:1999yu} and is   summarized in Table \ref{T4}.
\begin{table}[H]
\be
\def\arraystretch{1.3}
\begin{array}{|cccc|}
\hline
 & SO(6) & SU(2)_{R} & \Delta \\
 \hline
 \varphi & (0,0,0) & \mathbf{p+1} & 2\,p \\
 \psi^{+} & (\tfrac{1}{2}, \tfrac{1}{2}, \tfrac{1}{2}) &  \mathbf{p}   & 2\,p+\tfrac{1}{2} \\
 V_{m} & (1,0,0) & \mathbf{p-1} & 2\,p+1 \\
  \psi^{-} & (\tfrac{1}{2}, \tfrac{1}{2}, -\tfrac{1}{2}) &  \mathbf{p-2} & 2\,p+\tfrac{3}{2} \\
 \varphi' & (0,0,0) & \mathbf{p-3} & 2\,p +2\\
 \hline
\end{array}\notag
\ee
\caption{Short multiplets $\SS^{(1,0)}_{p}$ of $OSp(2,6|2)$ that  appear in the 
product of $p$   copies of $(1,0)$ doubleton  (hypermultiplet). 
 }
\label{T4}
\end{table}

Here $\Delta$ is the scaling dimension
of the conformal group $SO(2,6)$ related to the canonical dimension of the corresponding 6d field $\Phi$ 
  by $\dim\Phi = 6-\Delta$. We indicated
   also  the $SU(2)$  R-symmetry representations.\foot{$\bf p$    stands for a 
   representation of dimension $p$; fields with negative $SU(2)_{R}$ dimensions  should  be  dropped.
   For a  field in  $p$-dimensional representation of $SU(2)_{R}$  the $d$ values of the $R$-charge
are  $
-\frac{p-1}{2} + k, \quad k = 0,..., p-1.
$
}
  $\psi^\pm$ are positive/negative  chirality MW   spinors while  $\vp$ and $\vp'$ are scalars. 
  The vector $V_m$ is non-gauge (``massive") one  for $p >2$   but   has the 
  standard   gauge invariance for $p=2$.
  
  The  $p=2$   case of $\SS^{(1,0)}_{p}$ is thus  $V^{(1,0)}$  in \rf{19}.
  Note also  that the formal      $p=1$ case   of   field   content in Table \ref{T4}  is the same  as 
  the scalar multiplet $S^{(1,0)}$ in \rf{18}  but with the opposite chirality 
  of  the  fermion, {\em i.e.} $\SS^{(1,0)}_{1}=S^{(0,1)}$. 
   Thus the chiral  anomalies of $\SS^{(1,0)}_{1}$
  will be opposite   in sign to the anomalies of $S^{(1,0)}$.
  In what follows we shall assume that $p >1$, as required by the actual construction of $\SS^{(1,0)}_{p}$
  as a tensor product.
  

 From the canonical   dimensions of the fields   one can determine the power of derivatives in 
 kinetic terms in the corresponding 6d Lagrangian
%
%
%
%
 \be \la{31} 
\begin{split}
\mc L=& \varphi\,\Box^{2p-3}\, \varphi+\overline\psi^{+}\slashed{\partial}^{4p-5}\psi^{+}
+V_{m}\,\Box^{2p-2}\,V_{m}+\overline\psi^{-}\slashed{\partial}^{4p-3}\psi^{-}+
\varphi'\,\Box^{2p-1}\,\varphi'  \ , 
\end{split}
\ee
where each field is assumed to  transform under $SU(2)_{R}$ according to representations
in Table~\ref{T4}.
\footnote{From the kinetic terms we can  compute the  corresponding numbers of  dynamical  d.o.f. 
for bosons and fermions ($p>2$)
$\qquad 
\nu_{b} = (p+1)(2p-3)+(p-3)(2p-1)+6\,(p-1)(2p-2) = 4(2p-3)(2p-1), $ 

\noindent  $ 
\nu_{f} = -2[p(4p-5)+(p-2)(4p-3)] = -4(2p-3)(2p-1),
$
so that  $\nu_{b}+\nu_{f}=0$.  The  above expressions are formally true also for $p=2$   when the vector is a 
gauge-invariant one but 
 there is no  second scalar $\vp'$   and  this  has the same effect as subtracting
the gauge mode of $V_{m}$. 
}

 Let us   first   compute   the corresponding chiral  $SU(2)_R$ and  gravitational anomalies. 
 The gravitational anomaly coefficients  $\g,\d$ in \rf{11}  here  get contributions from the chiral fermions
  $\psi^+$ and $\psi^-$   and  do not depend on powers of derivatives in the kinetic terms, {\em i.e.} 
  are the same as for the usual $\slashed{\partial} $ fermions. As we have $p$  of $\psi^+$  fermions 
  and $p-2$  of $\psi^-$  fermions the total   contribution  is as of $p - (p-2) =2$   positive-chirality fermions, {\em i.e.}  it 
  is the same as for the  standard vector  multiplet  or  for $V^{(1,0)}$ in \rf{19} (see Table \ref{T2}). 
  Thus  the $\g,\d$  coefficients 
    are $p$-independent
  \be\la{32}
\SS^{(1,0)}_{p}:\qquad \qquad  \gamma \te = -\frac{7}{240},\qquad  \qquad \delta = \frac{1}{60} \ .
\ee
 The coefficients $\a$ and $\b$ in \rf{11}  are proportional to the sums 
 of  second and fourth powers of the R-charges.\footnote{Here we consider the Cartan 
  $U(1)$ subgroup of the $SU(2)_{R} $ as  
the  4th order anomaly $c_{2}^{2}$  is  determined  by  the sum of fourth powers of the $J_{z}$
eigenvalue, see, {\em e.g.},  \cite{Kumar:2010am}.}
 Taking again into account that  the chiral anomaly   receives only  the 
 contributions of the two opposite  chirality  fermions  in Table \ref{T4},     we get 
\be\la{33}
\begin{split}
\sum
r^{2} &= \sum_{k=0}^{p-1}{\te \left(-\frac{p-1}{2}+k\right)^{2}}-
\sum_{k=0}^{p-3}{\te \left(-\frac{p-3}{2}+k\right)^{2}} =\te  \frac{1}{2}(p-1)^{2}, \\
\sum
r^{4} &= \sum_{k=0}^{p-1}{\te \left(-\frac{p-1}{2}+k\right)^{4}}-
\sum_{k=0}^{p-3}{\te \left(-\frac{p-3}{2}+k\right)^{4}} = \te \frac{1}{8}(p-1)^{4}.
\end{split}
\ee
Including the normalization constants   so that  for $p=2$ we  recover the vector multiplet values  in Table \ref{T2},
 we get 
\be
\label{34}
\SS^{(1,0)}_{p}: \qquad\qquad  \alpha =  -(p-1)^{4}\ ,\qquad \qquad 
 \beta = -\tfrac{1}{2}(p-1)^{2} \ . 
\ee
As expected,   for $p=1$  the expressions for the chiral anomalies  in \rf{32}  and \rf{34} 
are the same as for $S^{(1,0)}$ in Table \ref{T2} up to  an opposite  overall sign.

Next, let us  turn to the conformal anomalies. 
To compute  the  conformal anomaly a-coefficient   corresponding to $\SS^{(1,0)}_{p}$
 we may use  the 
general expression for $\aa$ of a field  associated to a representation $(\Delta; h_1,h_2,h_3)$
of the   conformal group $SO(2,6)$   given in \ci{Beccaria:2014qea,Beccaria:2015uta}:\foot{
Here ${\bf{h}}=(h_1,h_2,h_3)$, \ $\bh= h_1 + h_2 + h_3$  and the 
dimension $\dd(\mathbf{h})$  of the $SO(6)$   representation $\mathbf{h}$ is 

$
\dd(\mathbf{h}) =\frac{1}{12}(1+h_{1}-h_{2})(1+h_{2}-h_{3})(1+h_{2}+h_{3})
(2+h_{1}-h_{3}) (2+h_{1}+h_{3})(2+h_{1}+h_{2}) $.

Note that 
$$
{\del \aa(\Delta;{\bf{h}})  \ov \del \Delta}= {\te \frac{(-1)^{2\bh} \dd(\bfh)}{96\times360}}
(\Delta - h_1 + 5) (\Delta + h_1 -1)  (\Delta - h_2 -4 )  (\Delta + h_2 -2) 
 (\Delta - h_3 -3) (\Delta + h_3 -3)\ . 
$$
}
\ba\no
& \aa(\Delta;{\bf{h}}) = -\frac{(-1)^{2\bh} \dd(\bfh)}{96\times37800} \, (\Delta -3)
\Big[15 (\Delta -3)^6 \\
&  \qquad \qquad \qquad   -21 (\Delta -3)^4 \left[h_3^2+h_1 \left(h_1+4\right)+h_2 \left(h_2+2\right)+5\right]\no 
\\ \no 
&   \qquad \qquad \qquad 
 +  35 (\Delta -3)^2 \big[\left(h_1+2\right)^2
   \left(h_2+1\right)^2+\big(h_1 \left(h_1+4\right)+h_2 \left(h_2+2\right)+5\big) h_3^2\big] \\
   &    \qquad \qquad \qquad   
   - 105  \left(h_1+2\right)^2
   \left(h_2+1\right)^2 h_3^2\Big] \ . \la{35}
\end{align}
Summing up  contributions   of all  fields  with their multiplicities  in Table \ref{T4}, we get\foot{Explicitly, 
the  separate contributions  $\aa(\Delta; h_1,h_2,h_3)$ are 
\be\notag
\begin{split}
&\aa(2p; 0,0,0) =\te  -\frac{1}{1890}(p-1)^{7}+\frac{1}{540}\,(p-1)^{6}-\frac{1}{540}\,(p-1)^{5} \te 
+\frac{11}{12960}\,(p-1)^{3}-\frac{1}{2880}\,(p-1)^{2}+\frac{1}{72576}, \\
&\aa(2p+\tfrac{1}{2}; \tfrac{1}{2},\tfrac{1}{2},-\tfrac{1}{2}) 
\te = \frac{2}{945}\,(p-1)^{7}-\frac{1}{270}\,
(p-1)^{6}-\frac{1}{270}\,(p-1)^{5}\te +\frac{1}{144}\,(p-1)^{4}
+\frac{1}{810}\,(p-1)^{3}\\ &\qquad \qquad \qquad \qquad \te 
-\frac{1}{360}\,(p-1)^{2}+\frac{191}{1451520}, \\
&\aa(2p+1; 1,0,0) = \te   -\frac{1}{315}\,(p-1)^{7}+\frac{1}{90}\,(p-1)^{5}-\frac{1}{240}\,(p-1)^{3}, \\
&\aa(2p+\tfrac{3}{2}; \tfrac{1}{2},\tfrac{1}{2},\tfrac{1}{2}) =\te 
\frac{2}{945}\,(p-1)^{7}+\frac{1}{270}\,(p-1)^{6}-\frac{1}{270}\,(p-1)^{5}-\frac{1}{144}\,(p-1)^{4} +\frac{1}{810}\,(p-1)^{3}\\ &\te 
 \qquad \qquad \qquad \qquad+\frac{1}{360}\,(p-1)^{2}-\frac{191}{1451520}, \\
&\aa(2p+2; 0,0,0) =\te -\frac{1}{1890}\,(p-1)^{7}-\frac{1}{540}\,(p-1)^{6}-\frac{1}{540}\,(p-1)^{5}  
+\frac{11}{12960}\,(p-1)^{3}+\frac{1}{2880}\,(p-1)^{2}-\frac{1}{72576}.
\end{split}
\ee
}
\be
\label{36}
\SS^{(1,0)}_{p}: \qquad \qquad  \aa =\te 
 \frac{1}{72}\,(p-1)^{4}-\frac{1}{144}\,(p-1)^{2}+\frac{11}{34560} \ . 
\ee
For $p=2$  this gives the value for $V^{(1,0)}$ in \rf{20}.\foot{The expression 
\rf{36} was derived   assuming $p >1$; still, if we formally set 
$p=1$  we get  the value  opposite
to the  value of $S^{(1,0)}$ in Table \ref{T1}  while the conformal anomaly of 
 $\SS^{(1,0)}_{1}=S^{(0,1)}$ (cf. Table \ref{T4})   should be the same as   of $S^{(1,0)}$.}

It  is  straightforward  also to directly 
compute the conformal  anomaly on a Ricci  flat background  by assuming 
that   Weyl-covariant   kinetic operators   appearing   in the generalization  of the   Lagrangian \rf{31} 
to  curved background factorize   into a product of  the standard 2-derivative  operators (cf.  Appendix \ref{A}). 
We then find 
\ba
\SS^{(1,0)}_{p}: \ \     A_6\big|_{R_{mn}=0}   &= \big[(p+1)\,(2p-3)+(p-3)\,(2p-1)\big]\, A_6(\varphi) +(p-1)\,(2p-2)\, A_6(V) \no \\
&\ \ \ \ + \big[p\,(4p-5)+(p-2)\,(4p-3)\big]\, A_6(\psi)  =\te -\frac{1}{11520}\, E_{6} \ . \la{37}
\end{align}
Here $A_6(\varphi)$, $A_6(V)  $  and $A_6(\psi) $  are  conformal anomalies 
for   standard Laplacians   defined on scalars, vectors and spinors  on a Ricci flat  background. 
This expression  was  derived   for $p >2$ but is also valid for $p=2$, {\em i.e.} for $V^{(1,0)}$   when the vector is a gauge-invariant  one. 
 
 Comparing \rf{37}   to the general expression  in \rf{6}   and using  the expression for $\aa$ in \rf{36} 
 we conclude that in addition to the expected relation \rf{7}, {\em i.e.}   $\cc_3= {1 \ov 6} (\cc_{1}-2\cc_{2})$, 
   we get 
 \be 
\SS^{(1,0)}_{p}: \qquad\qquad   \cc_1 + 4 \cc_{2} =192 \aa  -  \te \frac{1}{60}= 
{ 8 \ov 3} (p-1)^4 - { 4 \ov 3} (p-1)^4  + { 2 \ov 45}  
 \ . \la{38} 
\ee
Using the  results for the chiral anomalies  in \rf{32}, \rf{34}   we conclude that the value of 
the a-coefficient in \rf{36} is indeed consistent with the general expression \rf{12}  of \ci{Cordova:2015fha}. 
Our suggested  expressions \rf{14}  for $\cc_1$ and $\cc_2$   give 
\be \la{39}
\begin{split}
\te \cc_{1} = \frac{4}{3}\,(p-1)^{4}-\frac{2}{3}\,(p-1)^{2}+\frac{1}{27}\ , \qquad \qquad  \cc_{2} =\frac{1}{3}\,(p-1)^{4}-\frac{1}{6}\,(p-1)^{2}+\frac{1}{540}  \ . 
\end{split}
\ee
These   are indeed consistent with  \rf{38}, thus   providing an additional test of  eq. \rf{14}. 
Let us  note  also  that  $\cc_{1}-4\,\cc_{2} = \frac{4}{135}$  is independent of $p$, 
{\em i.e.} is the same as for $V^{(1,0)}$.\foot{Once again,  
for $p=1$  the $\cc_i$  in \rf{39}   are opposite to the values  corresponding to 
$S^{(1,0)}$ in Table \ref{T1}   which is due to the fact that 
we applied  the relations  between chiral and conformal anomalies 
valid for $(1,0)$   multiplets  while the $p=1$ example   is  formally a 
$(0,1)$  scalar multiplet (cf. footnote above eq. \rf{12}).}

\section{Higher derivative   spin 2   superconformal  multiplets}

In this section  we shall    consider  the   chiral and conformal   anomalies 
  of   higher  derivative  6d  superconformal multiplets generalizing $(2,0)$
     conformal  supergravity (CSG)  multiplet. 

Let us    start with  the 
Kaluza-Klein
  spectrum  \cite{Casher:1984ym,Gunaydin:1984wc,vanNieuwenhuizen:1984iz} 
of 11d   supergravity compactified on $S^4$  
 given 
in Table \ref{T5} (see also \cite{D'Hoker:2000vb}).
\begin{table}[H]
\be
\begin{array}{|c|l|c|}
\hline
& (\Delta; h_{1},h_{2},h_{3}) & USp(4) \\
\hline
&(2 p;0,0,0) & [0,p] \\
&(2 p+\frac{1}{2};\frac{1}{2},\frac{1}{2},\frac{1}{2}) & [1,p-1] \\
&(2 p+1; 1,1,1) & [0,p-1] \\
p\ge 2&(2 p+1; 1,0,0) & [2,p-2] \\
&(2 p+\frac{3}{2};\frac{3}{2},\frac{1}{2},\frac{1}{2}) & [1,p-2] \\
&(2 p+2;2,0,0) & [0,p-2] \\
\hline
\hline
&(2 p+\frac{3}{2};\frac{1}{2},\frac{1}{2},-\frac{1}{2}) & [3,p-3] \\
p\ge 3 &(2 p+2;1,1,0) & [2,p-3] \\
&(2 p+\frac{5}{2}\frac{3}{2},\frac{1}{2},-\frac{1}{2}) & [1,p-3] \\
&(2 p+3;1,1,-1) & [0,p-3] \\
\hline
\end{array}
\ \ \ 
\begin{array}{|c|l|c|}
\hline
& (\Delta; h_{1},h_{2},h_{3}) & USp(4) \\
\hline
&& \\
&& \\
&(2 p+2;0,0,0) & [4,p-4] \\
&(2 p+\frac{5}{2};\frac{1}{2},\frac{1}{2},\frac{1}{2}) & [3,p-4] \\
p\ge 4 &(2 p+3;1,0,0) & [2,p-4] \\
&(2 p+\frac{7}{2};\frac{1}{2},\frac{1}{2},-\frac{1}{2}) & [1,p-4] \\
&(2 p+4;0,0,0) & [0,p-4] \\
&& \\
&& \\
&& \\
\hline 
\end{array}
\nonumber
\ee
\caption{$SO(2,6)\times USp(4)$ representations of  fields  of 11d supergravity on 
AdS$_{7}\times S^{4}$ vacuum. Each level $p$ corresponds to 6d $(2,0)$   superconformal multiplet. The canonical dimension of the corresponding 6d fields is 
$\Delta_-= 6 - \Delta$.
} 
\label{T5}
\end{table}
 The massless level $p=2$  is represented by the fields of  maximal gauged  7d supergravity 
with AdS$_{7}$  vacuum.
 The corresponding  6d   ``massless''  $OSp(2,6|4)$ superconformal multiplet 
  is that of $(2,0)$  conformal supergravity 
  \ci{Bergshoeff:1999db,Beccaria:2015uta}.\foot{In Table \ref{T5} we have chosen chirality assignments 
  so that   for $p=2$    they correspond to the 
  canonical choice  in the $(2,0)$  conformal supergravity   \ci{Bergshoeff:1999db}, {\em i.e.}
  the  fermions and gravitino have positive chirality and  the antisymmetric tensor is self-dual.}
   Similarly,   the  7d  multiplet  formed by fields   belonging to $p>2$   level
corresponds to   higher derivative  ``massive"   $(2,0)$    superconformal    multiplet
 in 6d (which we shall denote  as  CSG$_p$).\foot{Fields  of  ``massive"  conformal   group 
representations  do not have  gauge invariances   that   are  present  in the ``massless"   case.} 
Equivalently,   just like    the  higher derivative vector multiplets  in Table \ref{T4} 
appear  in the product of $p$ copies of $(1,0)$   hypermultiplet  $S^{(1,0)}$, 
the \csg   multiplets appear  in the product of $p$   copies of $(2,0)$  tensor multiplet $T^{(2,0)}$
\cite{Gunaydin:1984wc}.  

 Below  we shall     compute the  chiral (gravitational and R-symmetry)  anomalies  and  the 
conformal  anomaly $\cc$-coefficient   in \rf{8}   of these $(2,0)$    \csg  multiplets 
complementing the result for their  conformal anomaly 
$\aa$-coefficient found in \ci{Beccaria:2014qea,Beccaria:2015uta}.
The anomaly   coefficients    will  all be proportional   to the same 
 factor   $ 6 p (p-1) +1$   and  will  satisfy the expected relations \rf{16}, \rf{17}, \rf{77}. 
In particular, we   will confirm the conjecture of \ci{Beccaria:2015uta}   that  the system 
of  $(2,0)$  conformal   supergravity  coupled to 26 tensor  multiplets is  completely anomaly  free.


\subsection{Chiral anomaly coefficients} 

 The  fields   in Table \ref{T5} contributing to  chiral (gravitational and R-symmetry)  anomalies in \rf{11}  are the 
MW fermions  $\psi^\pm\sim ( \ha,\ha, \pm \ha)$, MW  conformal gravitini $\psi^\pm_m \sim 
(\te {3\ov 2},\ha, \pm \ha)$  and 
(anti)self-dual  rank 3  antisymmetric tensors $A^\pm_{mnk} \sim (1,1,\pm 1)$. 
Contributions  of   these  fields  are to   be summed up with  multiplicities corresponding to their 
$USp(4)= SO(5)$  R-symmetry  representations.\footnote{The dimension of the  $USp(4)$   representation $[a,b]$  
 ($a,b$   are Dynkin labels) is 
 
$\dim(a,b) = \frac{1}{6}\,(a+1)(b+1)(a+b+2)(a+2b+3)$.} 

To find $USp(4)$ multiplicity  of a particular conformal  field  at level $p$  one  is to  add up 
dimensions of  the corresponding $USp(4)$ representations in Table \ref{T5}.
Since  the positive-chirality fields   contribute  to chiral anomalies with the opposite sign 
compared to the  negative-chirality ones, we may  express   the total result  in  terms of the 
effective  numbers of,  {\em e.g.},    the positive-chirality  fields.
 Counting  the  negative chirality   fermions 
as minus  the  positive  chirality ones    we find for the  effective number 
of  the  positive  chirality   MW spinors $\psi^+ \sim (\tfrac{1}{2}, \tfrac{1}{2}, \tfrac{1}{2})$  at level $p$:  
\be 
n(\psi^+)=\dim(1,p-1)-\dim(3,p-3) +\dim(3,p-4)-\dim(1,p-4) = 2p(p-1)+8 \no \  .  \ee
Similarly, for the effective number of positive chirality  MW
gravitini    we find 
$n(\psi^+_m) = 2p(p-1)$,  while 
the effective  number of self-dual 3-form fields is   $n(A^+_{mnk}) =2p(p-1)+1$. 

The  conformal  6d  fields  corresponding to representations in Table \ref{T5} 
are   non-standard  having  higher-derivative kinetic terms 
 (with the number of derivatives determined by  canonical dimension $6-\Delta$).
 In general, the  chiral   anomalies of higher-derivative  fermions    will be same as  
 anomalies of their lowest-derivative counterparts.  For 
 the gravitino  $\psi^+_m$ and the antisymmetric tensor $A^+_{mnk} $
at  levels  $p >2$ we will have an additional  complication:  they will   be 
 conformal   and  ``massive" ({\em i.e.} will not have   usual gauge invariance). 

Let us  start with the gravitational anomalies   and 
 first recall the expressions  \cite{AlvarezGaume:1983ig} for the 
  purely gravitational parts of the 6d anomaly polynomial  $\mc I_{8}$ in \rf{11} 
for the  positive  chirality   MW fermion,  
 the 
standard  gauge-invariant  (real)  self-dual   rank 3 antisymmetric tensor ($A\equiv H= d T$, 
with potential denoted as $T^+$) 
and the standard (1st-derivative, gauge-invariant)   positive   chirality MW gravitino\foot{Using 
the field content \rf{18},\rf{19}  of the   $(1,0)$   scalar, tensor  and vector multiplets  
one can check  that the values  of $\g,\d$   in Table \ref{T2} 
indeed follow   from   \rf{40}.}
\be
\label{40}
\begin{split}
\te 
&\te \mc I_{8}({{1\ov 2}^+})=
\te  - \frac{1}{16 \times 6!}(7\,\rp_{1}^{2}-4\,\rp_{2}), \qquad
\qquad 
\mc I_{8}({T^+})
=\te - \frac{1}{16\times 6!}(32\,\rp_{1}^{2}-224\,\rp_{2})\ , 
 \\
& \qquad \qquad \qquad \qquad 
\mc  I_{8}({\te {3\ov 2}^+})= 
\te - \frac{1}{16 \times 6!}(275\,\rp_{1}^{2}-980\,\rp_{2})\ .
\end{split}
\ee
Given  a generic rank 3   antisymmetric tensor  (with  $20 = \binom{6}{3}$
components) we may 
represent it in terms of  two independent 
transverse 2-tensors  $T_{mn},\ \widetilde T_{mn}$
with  $10+10$ components as\footnote{This is analogous to the  representation 
of  the antisymmetric rank 2  tensors  in 4d conformal   supergravity  used 
in section 2.3 of \cite{Fradkin:1985am}, see also a discussion in \ci{Carrasco:2013ypa}.} 
\be\la{41} 
A_{mnk} = \partial_{[m} T_{nk]}+ \epsilon_{mnklpq}\,
\partial_{[l} \widetilde T_{pq]} \ .
\ee
Here only  the  transverse parts of $T$ and $\td T$ contribute, {\em i.e.}
$A$  is thus  expressed  in terms of   two standard gauge-invariant  3-form field  strengths 
$H=dT$ and $\widetilde H = d \widetilde T$. 
Similarly, the contribution of a self-dual $A$  to chiral anomalies 
  will be equivalent to the   contributions 
of self-dual parts of $H$ and $\widetilde H$, {\em i.e.}  it will be  twice   the  standard 
antisymmetric tensor  contribution  in \rf{40} (or, equivalently,  the contribution of one complex 
$T^+$ field). 
%
%
%

The gravitino  $\psi^\pm_m \sim  ( \te {3\ov 2},\ha, \pm \ha)$   in Table \ref{T5}
 is different from the  standard  gravitino  discussed in  \cite{AlvarezGaume:1983ig} 
 in  two respects (in addition to  having 
  higher-derivative kinetic term):  (i) for any $p$  it is   conformal ({\em i.e.} its  $\gamma$-trace is zero);
 (ii) for $p >2$  it  has no   gauge invariance. 
 Thus compared to the standard negative-chirality gravitino    anomaly  
 $\mc I_{8}({{3\ov 2}^+})$  in \rf{40}  the anomaly of  the massive conformal  $\psi^+_m$ 
   should have   the 
  negative-chirality fermion ($\gamma$-trace)  anomaly  $\mc I_{8}({{1\ov 2}^-})= - \mc I_{8}({{1\ov 2}^+})$
   subtracted  and the 
  positive-chirality fermion ($\del_m \epsilon$  gauge degree of freedom)  anomaly $\mc I_{8}({{1\ov 2}^+})$
  added, or, equivalently, the  ghost contribution should not  be subtracted.
  Thus   for $p\ge 2$ we  should   have 
\be
\begin{split}
\mc I_{8}(\psi^+_m) 
  \te =\mc I_{8}({ {3\ov 2}^+} )
  + 2 \,\mc I_{8}({{1\ov 2}^+} ) \ . \la{42} 
\end{split}
\ee
As a result, the total gravitational anomaly  polynomial  of \csg  multiplet is found to be 
\be
\label{86}
\begin{split}
\mc I_{8} (\text{CSG}_p) &=\te  [2p(p-1)+8] \, \mc I_{8}({{1\ov2}^+})
+ 2 \, [2p(p-1)+1]\, \mc I_{8}({T^+} )
\\
&  \qquad  +   2p(p-1) \, \te  \big[\mc I_{8}({{3\ov2}^-})  +  2 \,\mc I_{8}({{1\ov 2}^+}) \big]
\\
&=\te  -\frac{1}{96}[6p(p-1)+1]\,(\rp_{1}^{2}-4\rp_{2}) \ .
\end{split}
\ee
The corresponding   gravitational anomaly coefficients in \rf{11} are thus 
\be
\label{88} 
\text{CSG}_{p}: \ \ \ \ \ \  
\te \gamma = -\frac{1}{4}\,\delta = -{1\ov 4} \big[ 
  6p(p-1)+1\big]\  . 
\ee
This  is in agreement with the relation \rf{16} expected for $(2,0)$   multiplets. 

The  expressions \rf{86}, \rf{88}   were     derived for $p >2$ (when gravitino is massive)  
but are  actually valid also for $p=2$: in this case  the  gauge-invariant conformal  gravitino 
contribution is $\mc I_{8}({{3\ov2}^+})  +   \,\mc I_{8}({{1\ov 2}^+})$
but   we  formally get also    ``extra"  4 chiral fermion contributions  from the 
multiplicity factor  (see  Appendix \ref{C.2}).

%
%
%
%


Similar results are found  for the R-symmetry and mixed anomaly  coefficients 
$\a$ and $\b$  in \rf{11}  (see Appendix \ref{C}):
\be \la{43}
\text{CSG}_{p}: \ \ \ \ \ \qquad    \a= 2 \b = - 2\big[ 6 p (p-1) + 1 \big] \ . 
\ee
We thus   observe    that   for any $p$ all  4 chiral anomaly   coefficients 
$\vec\alpha = (\a,\b,\g,\d)$  are  proportional to the  chiral anomaly coefficients  of  the $T^{(2,0)}$ 
multiplet in Table \ref{T2}, {\em i.e.} 
\be\la{44}
\vec\alpha(\text{CSG}_{p}) = -2\big[6p(p-1)+1\big]\,\vec\alpha(T^{(2,0)})\ , \qquad \qquad 
\te \vec\alpha(T^{(2,0)}) = (1,\frac{1}{2},\frac{1}{8},-\frac{1}{2}) \ . 
\ee
In particular, for $p=2$   when \csg is the multiplet of $(2,0)$   conformal supergravity,
we get  
\be\la{45} 
\vec\alpha \big(\text{CSG}^{(2,0)}\big) +26\, \vec\alpha \big(T^{(2,0)}\big) = 0 \ .
\ee
As a result, the gravitational  and R-symmetry anomalies 
   of $(2,0)$  conformal supergravity   can be cancelled by adding 
    26   tensor  multiplets $T^{(2,0)}$.

\subsection{Conformal anomaly coefficients} 

The conformal anomaly 
$\aa$-coefficient  for \csg   multiplet  was found in \ci{Beccaria:2014qea,Beccaria:2015uta}
by a computation  on $S^6$   
\be \la{399}
\aa(\CSG_p) = - 2 \big[ 6 p (p-1) +1\big] \, \aa(T^{(2,0)})   \ , \te  \qquad\qquad  \qquad  \aa(T^{(2,0)})   = - {7 \ov 1152} \ ,
\ee
where $ \aa(T^{(2,0)})  $ is the $(2,0)$ tensor multiplet value in \rf{13}. 
Comparing  the expression  \rf{399}   with the chiral anomaly  result  \rf{44} 
we conclude that  the $(1,0)$  relation \rf{12}  of \ci{Cordova:2015fha}
or the  relation for $\aa$-coefficient    in \rf{17}   is indeed satisfied.

While  a direct computation   of the $\cc_i$    anomaly   coefficients in \rf{4}  for  the \csg  multiplet
may be feasible 
 by assuming  factorization of all   kinetic operators on a Ricci flat background, 
 here 
 we  shall    apply the  expected   relations \rf{14}  or \rf{17}  between   chiral and conformal 
 anomalies of superconformal multiplets 
 implying  in  view of  \rf{44} that $\cc_i$   in \rf{4}  are all proportional to $\cc$ as  in \rf{8}   where 
 $\cc$ is given  by \rf{17},\rf{77}
 \be 
 \la{46}
\cc =\aa - \te { 1 \ov 192} \delta = -2\big[6p(p-1)+1\big]\, \cc(T^{(2,0)}) \ , \ \ \ \ \qquad\qquad   
\te \cc(T^{2,0}) =  - { 1\ov 288} \ . \ee
As a result,  applying this to the  $p=2$  case we conclude  that   all  chiral and conformal  anomalies 
of (2,0)    system of 6d  conformal  supergravity  plus 26 tensor multiplets 
vanish.

\section{Some   interacting  $(1,0)$    superconformal   theories }
Let us now  consider   some  examples of interacting $(1,0)$ superconformal theories. 

First,  let us   summarize   the expressions  for the  chiral, gravitational \ci{Harvey:1998bx,Freed:1998tg},  
and conformal \ci{Henningson:1998gx,Tseytlin:2000sf,Beem:2014kka,Beccaria:2014qea,Ohmori:2014kda} 
anomaly coefficients of interacting $(2,0)$  $A_N$  theory describing $N$ coincident 
  M5 branes:
\ba
\la{21} 
&(2,0): \ \ \ \   \qquad   \a= N^{3}-1\ , \qquad\qquad  \b= 4 \g = - \delta = \tfrac{1}{2}(N-1) \ ,\\
&\te   \aa= -\frac{1}{288}\, (4 N^{3}- \frac{9}{4}\,N- \frac{7}{4})=  -\frac{1}{288}\, (N-1)\big[( 2N+1)^2 + { 3\ov 4}\big]  \ , \ \ \ \ \ \la{22} \\
&\te 
\cc = -\frac{1}{288}\, (4 N^{3}-3 N -1)=   -\frac{1}{288}\, (N-1) ( 2N+1)^2\ , \qquad   \la{23}
 \cc_1 =4 \cc_2 = -12 \cc_3 =96 \cc\ .
   \end{align}
These   values  are  perfectly consistent with \rf{16}--\rf{77}. 
The leading order $N^3$ terms in \rf{22},\rf{23}  follow   \ci{Henningson:1998gx}
from  7d supergravity  $(R + \Lambda) $ terms  found  upon compactification of 11d theory on $S^4$, 
the subleading  order $N$ terms   originate from $R^4$   corrections in 11d action \ci{Tseytlin:2000sf}
and order $N^0$ terms  are  reproduced by 1-loop 11d  supergravity corrections \ci{Beccaria:2014qea,Mansfield:2003bg}.

There are also  two  cases of interacting $(1,0)$  superconformal 
theories  with known chiral anomalies. 
The first is 
 the $\mc E_{N}$  theory \ci{Ganor:1996mu}  on the world-volume of $N$ small coincident $E_{8}$
 instantons in the heterotic string  (E-string), or, equivalently, 
 the theory on $N$   M5-branes  on Horava-Witten $E_8$  9-brane. 
  The second  is the  $\mc T_{N,\Gamma}$ theory 
 describing  $N$  M5-branes on the orbifold singularity 
 $\mathbb{C}^{2}/\Gamma$ where $\Gamma$
 is a discrete subgroup of $SU(2)$ (see \ci{Ohmori:2014pca,Ohmori:2014kda}). 
 The corresponding  coefficients in the anomaly polynomial \rf{11} are given in Table \ref{T3}.
\begin{table}[H]
\be
\def\arraystretch{1.3}
\begin{array}{|c|c|c|c|c|}
\hline
 & \alpha & \beta & \gamma & \delta \\
 \hline
 \mc E_{N} & N(4N^{2}+6N+3) & -\frac{N}{2}(6N+5) & \frac{7N}{8} & -\frac{N}{2} \\
 \hline
\mc T_{N, \Gamma} & |\Gamma|^{2}N^{3}-2N|\Gamma|(r_{\Gamma}+1) & 
 N-\frac{N}{2}|\Gamma|(r_{\Gamma}+1)+\frac{d_{\Gamma}}{2} & 
 \frac{N}{8}+\frac{7d_{\Gamma}}{240} & -\frac{N}{2}-\frac{d_{\Gamma}}{60} \\
  & +2N+d_{\Gamma} & 
  &   &   \\
 \hline
\end{array}\notag
\ee
\caption{Chiral anomaly coefficients for the $\mc E_{N}$ and $\mc T_{N, \Gamma}$ 
$(1,0)$  theories. 
}
\label{T3}
\end{table}
Here $|\Gamma|$  is the order  of the discrete group $\Gamma$ 
and $r_\Gamma$ and $d_\Gamma$  are the rank and dimension of the associated Lie algebra  $G_\Gamma$  ({\em e.g.},   $G_\Gamma=SU(k)$  for $\G= \mb Z_k$). 

It follows from \rf{13}  that  the corresponding $\aa$-anomaly coefficients are 
\ci{Cordova:2015fha}
\ba&\aa(\mc E_{N}) = \te 
\frac{16 }{7} \Big(4N^3 +9 \,
   N^2+\frac{99}{16}\,N\Big)\,\aa_{\TT}  \  , \la{24}\\
\te &\aa(\mathcal T_{N, \Gamma})\te 
 =\frac{16}{7} \Big(
|\Gamma|^{2}N^{3} -\frac{3}{2} \big[ |\Gamma|(r_{\Gamma}+1) - \frac{5}{8}\big] N+\frac{251}{480}d_{\Gamma}
\Big)\,\aa_{\TT}\ . \la{25}
\end{align}
From  the expressions \rf{14}   with the  chiral  anomaly coefficients  from  Table \ref{T3} we find 
\ba
\te 
&\cc_{1}(\mc E_{N}) \te = -\frac{16}{3}\,N^{3} - {12} N^2  
- {76\ov 9}   N 
\ , \qquad \qquad \cc_{2}(\mc E_{N}) \te = -\frac{4}{3}\,N^{3}
- 3 N^2  
- {17\ov 9} N\ , \la{26} 
\\ 
&\cc_{1}(\mc T_{N, \Gamma}) \te = -\frac{4}{3}\,|\Gamma|^{2}\,N^{3}+\big[
2 \,|\Gamma|\,(r_{\Gamma}+1)-1  \big]\,N  -\frac{19}{27} \, d_{\Gamma},\no  \\
&\cc_{2}(\mc T_{N, \Gamma}) \te = -\frac{1}{3}\,|\Gamma|^{2}\,N^{3}+{1 \ov 4}  \big[
{ 2} \,|\Gamma|\,(r_{\Gamma}+1)
 - { 1} \big]\,N
-{91\ov 540} \, d_{\Gamma} \ . \la{27} 
\end{align}
Here  the leading large $N$  terms ($\sim N^{3}$) in $\cc_1$ and $\cc_2$  are in ratio 4:1
as   for a $(2,0)$ theory (cf. \rf{8}). This  is what should  be expected from the AdS/CFT as the $ N^{3}$
terms originate from the universal   Einstein term in  the dual 7d supergravity action. 

Furthermore,   a similar relation is true  also for the  first  subleading terms,  $\mc O(N^2)$ 
in $\mc E_{N}$  case and $\mc O(N)$ in $\mc T_{N, \Gamma}$ case, {\em i.e.} 
 \foot{The vanishing of the  two leading coefficients in \rf{28}  is  a consequence of the  special 
 choice of the parameter mentioned in  footnote \ref{new-foot}  that  we made in writing \rf{14}.
}
\be \la{28}\begin{split} \te 
&\cc_{1}(\mc E_{N})-4\cc_{2}(\mc E_{N}) 
= \te 0\cdot N^{3} +  0\cdot N^{2}-\frac{8}{9}\,N, \qquad \\
&\cc_{1}(\mc T_{N, \Gamma})-4\cc_{2}(\mc T_{N, \Gamma}) = \te 0\cdot N^{3} +  0\cdot N-\frac{4}{135}\,d_{\Gamma}\ .
\end{split}
\ee
This  fact should also  have a holographic explanation,  cf. \ci{HH},
  which  may  thus provide an  independent   check   of
 the relations \rf{14}. 

Another indirect support for our  relations  \rf{14} 
comes from  consideration of a   class of 6d $(1, 0)$ theories named {"very Higgsable"}  in 
\cite{Ohmori:2015pua}: they   admit a completely Higgsed branch where no tensor multiplets remain.
Compactifying these   theories  on $T^{2}$  one gets  4d  $\N=2$  superconformal theories 
with  the 4d conformal anomaly  coefficients $(\aa_{4\rm d}, \cc_{4\rm d})$
determined in terms of the coefficients in the 
 6d anomaly polynomial  $\mc I_{8}$  in \rf{11} as  \cite{Ohmori:2015pua}. In our notations, 
 \be
 \label{29}\te 
 \aa_{4\rm d} = \frac{1}{4!}(-12\,\beta+24\,\gamma-18\,\delta), \qquad\qquad 
 \cc_{4\rm d} = \frac{1}{4!}(-12\,\beta+64\,\gamma-8\,\delta)\ .
 \ee
 Using \rf{29}   together   with \rf{14}   we then find  the following  remarkable
 identity
  \foot{Again, the special   choice of the  parameter in footnote \ref{new-foot} leading to 
 the expressions in \rf{14}  is uniquely selected   if one insists on 
 reproducing the relation  \rf{30}.}
  \be\la{30} 
 \cc_{4\rm d}-\aa_{4\rm d} = -\tfrac{45}{32}(\cc_{1}- 4\cc_{2}) \ .
 \ee
 This  relates  the combinations  of  the conformal anomaly 
  coefficients 
  that vanish  in the  maximally  supersymmetric cases, {\em i.e.}  $\mc N=4$ in 4d and $(2,0)$  in 6d.


\section{6d conformal anomaly from 7d gravitational effective action} 

As already mentioned above, an   important   source of information  about conformal anomalies  of 
interacting  6d superconformal theories  is  AdS/CFT   \ci{Henningson:1998gx,Tseytlin:2000sf}. 
  One may  expect that   strong-coupling limit  of a 6d superconformal  theory is described   by some 
 effective  locally supersymmetric 7d theory with  an AdS$_7$ vacuum.  
 
 Such  7d   action may arise, {\em e.g.}, 
 from 11d   M-theory  effective   action     upon compactification  on some 4-space
 ($S^4$ in the case of the standard $(2,0)$ theory). 
 Ignoring for simplicity  all  other  fields  than the metric, 
 the effective 7d Lagrangian  will have the form 
 ${\mathcal L}= R + \Lambda +  R^2 + R^3 + R^4 + ...$   where  the expansion  in powers 
 of curvature  should correspond  to the strong-coupling (large $N$)  expansion in 
 the boundary  6d theory.  
 The  combinations  of the curvature  invariants 
should be such that they admit  supersymmetrization  consistent with the  amount 
 of supersymmetry of the boundary theory. 
 That  may  provide constraints \rf{7}, \rf{8} 
 on the  conformal anomaly $\cc_i$-coefficients   \ci{Kulaxizi:2009pz}
 that we   shall  discuss below.

\iffa 
leading to the relative weights of $\cc_{i}$ in the $(2,0)$ free tensor multiplet as well as in the 
interacting $\mathfrak{g}_{(2,0)}$ theories. \footnote{
This $(1,0)$ constraint  appears in the study of the angular dependence of the energy flux \cite{Hofman:2008ar}.
The factor $\cc_{1}-2\cc_{2}+6\cc_{3}$ multiplies a term allowed by conformal invariance with higher order
harmonic angular dependence. In superconformal models, this term is expected to be vanishing.}
\fi

\def \wa {u}  \def \wS {I}
\def \kk {{\rm k}} \def \a {\alpha} \def \LL {{\cal L}}

\subsection{Quadratic and cubic  curvature  corrections: linearized  approximation}

Let us   consider the 7d action including  the most general 
quadratic and cubic  curvature invariants\foot{We  shall  ignore terms with derivatives of 
the curvature as they will not contribute to the 
relevant  conformal anomaly coefficients.}
\ba
\label{11.2}
 &S = -\frac{1}{2\,\kappa_{7}^{2}}\,\int d^{7}x\,\sqrt{-g}\,\Big(R+\frac{30}{L^{2}}+\Delta  \LL  \Big) \ , \ \ \ \ \ \ \\
&\Delta\LL = 
L^{2}\,\sum_{i=1}^{3}\wa_{2,i} \wS_{2,i}+L^{4}\,\sum_{i=1}^{8}\wa_{3,i} \wS_{3,i}  \ . \label{1122}
\end{align} 
Here  $L$ is a  scale  (which to leading order is the same as the AdS$_7$ radius) 
  introduced to make   the coefficients $\wa_{n,i}$  dimensionless
   and $\wS_{r,i}$ are curvature contractions 
\be
\label{11.3}
\wS_{2,1} = R_{\m\n\l\r}^{2}, \qquad \wS_{2,2} = R_{\m\n}^{2}, \qquad \wS_{2,3} = R^{2},
\ee
\be
\label{11.4}
\begin{array}{cclcclccl}
\wS_{3,1} &=& R^{\m\n\l\r}R_{\r\r\k\a}R^{\k\a}{}_{\m\n}, & 
\wS_{3,2} &=& R^{\m\n}{}_{\l \r}R^{\l \k }{}_{\n \a}R^{\r \a}{}_{\m \k }, \  & 
\wS_{3,3} &=& R^{\m \l\r\k}R_{\r\k \l\a}R^{\a}{}_{\m }, \\
\wS_{3,4} &=& R\,R^{2}_{\m\n\l\r}, & 
\wS_{3,5} &=& R^{\m \n \r \l }R_{\r \m }R_{\l \n }, & 
\wS_{3,6} &=& R^{\m \n } R_{\n \l } R^{\l }{}_{\m },\\
\wS_{3,7} &=& R\,R^{2}_{\m\n}, & 
\wS_{3,8} &=& R^3. & 
 &&
\end{array}
\ee
The first  two terms   in \rf{11.2}    can be embedded into the  maximal 7d gauged  supergravity 
 so should describe  the $(2,0)$  superconformal theory  at the boundary, while the 
   higher   order  terms may or may not  (partially) break   supersymmetry. 

Starting   with \rf{11.2},   one may compute the coefficients of  the boundary conformal anomaly 
by  generalizing the  approach of  \ci{Henningson:1998gx}, {\em i.e.} extracting 
the IR logarithmic singularity in the action evaluated on a  classical solution with prescribed metric at the boundary.
Keeping only  terms linear in  coefficients $\wa_{r,k}$  in \rf{1122}\foot{Non-linear  corrections in the case of special $R^2$ and $R^3$   combinations  representing  Euler densities $E_4$ and $E_6$    were 
computed   in \ci{deBoer:2009pn,deBoer:2009gx,deBoer:2011wk,Hung:2011xb},  see Section 6.2  below.}
one finds for $\cc_i$  in \rf{4}  
 \cite{deBoer:2009pn,Kulaxizi:2009pz}\footnote{We thank  M. Kulahizi and 
 A. Parnachev for sending us a corrected version of the list of coefficients in \ci{Kulaxizi:2009pz}.}
\ba
\cc_{1} &= {\rm k}\,\big[-96+3072\,( \tfrac{5}{96}\,\wa_{2,1} + \tfrac{21}{32}\,\wa_{2,2}  +  \tfrac{147}{32}\,\wa_{2,3}\no  \\
& + \tfrac{9}{16}\,\wa_{3,1}+\tfrac{23}{16}\,\wa_{3,2}+\tfrac{5}{16}\,\wa_{3,3}-\tfrac{35}{16}\,\wa_{3,4}
   -\tfrac{63}{16}\,\wa_{3,5}-\tfrac{63}{16}\,\wa_{3,6}-\tfrac{441}{16}\,\wa_{3,7}-\tfrac{3087}{
   16}\,\wa_{3,8}) + \dots\big], \no \\
\cc_{2} &= {\rm k}\,\big[-24+ 3072\,(\tfrac{37}{384}\,\wa_{2,1} + \tfrac{21}{128} \,\wa_{2,2}  + \tfrac{147}{128} \,\wa_{2,3} \label{11.5}\\
& 
+\tfrac{9}{64}\,\wa_{3,1}+\tfrac{7}{64}\,\wa_{3,2}+\tfrac{37}{64}\,\wa_{3,3}-\tfrac{259}{64}\,\wa_{3,4}
   -\tfrac{63}{64}\,\wa_{3,5}-\tfrac{63}{64}\,\wa_{3,6}-\tfrac{441}{64}\,\wa_{3,7}-\tfrac{3087}
   {64}\,\wa_{3,8}
)+\dots\big],\no  \\
\cc_{3} &= {\rm k}\,\big[8+ 3072\,( \tfrac{3}{128}\,\wa_{2,1} - \tfrac{7}{128} \,\wa_{2,2}  - \tfrac{49}{128} \,\wa_{2,3}\no \\
& 
-\tfrac{41}{192}\,\wa_{3,1}-\tfrac{31}{192}\,\wa_{3,2}+\tfrac{9}{64}\,\wa_{3,3}-\tfrac{63}
   {64}\,\wa_{3,4}+\tfrac{21}{64}\,\wa_{3,5}+\tfrac{21}{64}\,\wa_{3,6}+\tfrac{147}{64}\,\wa_{3,7}+\tfrac{1029}
   {64}\,\wa_{3,8}
)+\dots\big]\ .  \no 
\end{align}
Here dots stand  for terms of higher order in $\wa_{r,i}$    and  the dimensionless factor   $\kk$  is 
\be\la{116}
{\rm k} = \frac{\pi^{3} L^5}{48\,\kappa_{7}^{2}}  \ . 
\ee
The leading order  ($\wa_{r,i}=0$)  values    \ci{Henningson:1998gx} are 
those of the large $N$ limit of the 
 $(2,0)$ theory    for which 
\be\la{117}
\frac{L^5}{2\kappa_{7}^{2}}= \frac{N^{3}}{3\,\pi^{3}},\qquad\qquad  {\rm i.e. }\qquad\qquad  {\rm k} = \frac{N^{3}}{72}\ , 
\ee
{\em i.e.}   $\cc_i$  are given by  the $N^3$   terms in \rf{23}   with $\cc= -\kk=  -\frac{N^{3}}{72}$
(which is  $4N^{3}$ times the $\cc$  anomaly of one free $(2,0)$ tensor multiplet \ci{Bastianelli:2000hi}). 
 
\iffa 
The interpretation is the following. 
There are 3 types of $RRR$  terms: those with index 
5,6,7,8   just  depend on Ricci tensor and renormalize the cosmological constant, so their coefficients 
satisfy $Q_2=0$.  The 3,4 structures renormalize  $R^2_{MNKL}$  so they   preserve $Q_{1}=0$, 
but break $Q_{2}=0$. A priori, the structures 1,2
break both constraints (the fact that $\wS_{1}$ does not enter $Q_2$ is just an accident, 
we may consider another combination where it will enter).
 {\bf   Here somehow it is assumed that $\wa_{2,i}$ and $\wa_{3,i}$   are of  the same order...  
so  this is a bit misleading and needs  clarification -- depends on context -- usually 
we have  curvature having its own scale  and its ratio to coeff scale $L$ is small like $\alpha' \over r^2$.
And we just do leading order perturbation theory in each type of coefficients.}
\fi 

To  find the $\aa$-coefficient in \rf{4} for the  theory \rf{11.2} 
one may follow  the  approach    used in 4d  case  in  \cite{Imbimbo:1999bj}  
and  compute   the gravitational effective action on the  corrected 
AdS$_7$ solution with  the  value of the    radius
that  extremises   the   action  evaluated on a test  AdS solution.\foot{The  boundary theory 
defined on $S^6$ and thus its 
conformal anomaly being determined just by the 
coefficient of the Euler density in \rf{4}.  The IR  divergent part of the  7d action  evaluated on modified AdS$_7$ solution 
(with log  IR divergence coming from the volume factor)
should then determine the   UV log part of the corresponding boundary effective action. 
AdS solution   will  always be a solution of the   modified gravitational equations  on symmetry grounds.} 
For the AdS$_7$  metric with radius $r$ we have 
\be\la{120}\te 
R_{\m\n\k\l} = -\frac{1}{r^{2}}(g_{\m\k}\,g_{\n\l}-g_{\m\l}\,g_{\n\k}), \qquad
R_{\m\n} = -\frac{6}{r^{2}}\,g_{\m\n},\qquad
R = -\frac{42}{r^{2}}.
\ee
The a-anomaly is then  proportional to the on-shell  value of the action  evaluated at the extremal  value $r_*$ of the 
 radius 
\ba\la{121} \te 
&\te\aa = -{\rm k}\, A(r_{*}), \qquad 
A(r) =  -\frac{1}{12\, L^5}r^{7}\,\big(-\frac{42}{r^{2}}+\frac{30}{L^{2}}\big) + O\big(\wa_{r,i}\big) \ , \qquad 
 \\
&
{dA\ov dr}\Big|_{r=r_{*}}=0 \ , \qquad \qquad  r_{*}=L  +   O\big( \wa_{r,i}\big)   \ , \qquad  A(r_*)=1 +   O\big( \wa_{r,i}\big)  \ , \la{121a}
\end{align}
where $\kk$ is given by \rf{116}. In $(2,0)$   theory  we then   find  to the  leading order 
$ \aa = -\frac{N^{3}}{72}$    \ci{Henningson:1998gx,Bastianelli:2000hi}    (cf. \rf{22}). 
Starting   with the general  $R^2 + R^3$  corrected action \rf{11.2},\rf{1122} 
  we get\foot{Note that if $\wa_{2,i}$ and $\wa_{3,i}$   are treated on an equal footing then   it is possible to  choose them    so that
the radius is   not modified   and yet the a-coefficient receives  a  correction:  
$
f_3 =  { 3 \ov 2}   f_2 \ , \ \ \       \aa= -\kk \big( 1  + 14 \,f_{2}\big)  
$.
Let us note also   that   in the   case when $R^2$ terms in \rf{1122} correspond 
to  the square of the Weyl tensor  (here $d=7$) 
$
C^{2}_{\m\n\l\r} = R_{\m\n\l\r}^2 - \frac{4}{d-2}\, R_{\m\n}^{2}+\frac{2}{(d-1)(d-2)}\,R^{2} 
$
we get  $f_2=0$ as expected:  in this case the AdS$_7$ solution is not modified   and the value  of the action is unchanged. 
The same   conclusion is found also for a combination of $C^2 + C^3$ terms. }
\ba
\label{1118}
&\te\aa= -\kk \big( 1-7\,f_{2}+14\,f_{3}\big)  \ ,   \qquad \qquad  r_{*} = L\,\big(1-\frac{3}{5}\,f_{2}+\frac{2}{5}\,f_{3}\big)\ , \\
\la{1117} 
&
f_{2} = \wa_{2,1}+3(\,\wa_{2,2}+7\,\wa_{2,3}),  \no \\
&f_{3} = \wa_{3,1}-\wa_{3,2}-3(\wa_{3,3}-7\wa_{3,4})+9(\,\wa_{3,5}+\,\wa_{3,6}+7\,\wa_{3,7}+49\,\wa_{3,8}).
\end{align}
\def \rr {{\rho}}
The    coefficients $\cc_i$ in \rf{11.5}  and $\aa$ in \rf{1118} can be written in a more compact form as 
\be
\label{588}
\begin{split}
\cc_{1} &= -96{\rm k}\,\big[1 -21   \overline \wa_{2,2}     -   \tfrac{5}{3}  \overline \wa_{2,1}   
- 2 (9\,\wa_{3,1}+  23\,\wa_{3,2}  ) + \dots\big], \\
\cc_{2} &= -24 {\rm k}\,\big[1  -21 \overline \wa_{2,2}    -  \tfrac{37}{3}\,\overline \wa_{2,1}     -2 (9   \,\wa_{3,1}  + 7 \,\wa_{3,2}  )
 +\dots\big], \\
\cc_{3} &=\ \ 8  {\rm k}\,\big[1   -21 \overline \wa_{2,2}        +9  \overline \wa_{2,1}  -   2( 41\,\wa_{3,1}  + 31 \,\wa_{3,2} )
+\dots\big], \\
 \aa &= -\kk \big[ 1 - 21  \overline \wa_{2,2}   -7 \overline  \wa_{2,1} 
  +14(\wa_{3,1}-\wa_{3,2})    + ...     \big]\ , 
\end{split}
\ee
where we introduced the combinations
\ba
& \overline \wa_{2,1}  \equiv   \wa_{2,1}  + 6  (\wa_{3,3}  -  7 \wa_{3,4}) \ , \la{rer} \ \ \ \ \   \\
&\overline \wa_{2,2}   \equiv   \wa_{2,2} + 7 \wa_{2,3}  -  6  (\wa_{3,5}+\,\wa_{3,6}+7\,\wa_{3,7}+49\,\wa_{3,8}) \ . \la{rer1}
\  
\end{align}
Thus    the 4  Weyl  anomaly coefficients   depend only on 4 non-trivial parameters.
This simplification   is due to the fact that 
 several  curvature invariants in \rf{11.4}    give  equivalent contributions   to the on-shell   action.
Indeed, $\overline \wa_{2,2} $ represents the contribution of $R^2 + R^3$ 
  terms   that depend  just on Ricci tensor and thus 
 renormalize   the  value   of the cosmological constant contribution   or the   overall scale of the 
 leading-order contribution to  the anomaly only. The coefficient
   $\overline \wa_{2,1} $  represents the terms that  reduce to $R^2_{\m\n\k\l}$ on the equations of motion. 
   This    ambiguity can be fixed   by setting some  redundant coefficients to zero. 
   For example, we may  
  demand 
   that the radius  is not modified, which requires according to   \rf{1118} that $3 f_2 = 2 f_3$. This 
   may be arranged  by fixing a combination of 
     parameters   that does not enter the conformal anomaly coefficients.
    
 \def \av {{\rm u}}\def \va {{\rm u}}
 
   Let us now   study  possible  supersymmetry   constraints. 
 Computing  the combinations $q_1$   in \rf{7}   and $q_2$  in \rf{8}   for the   coefficients in \rf{588}  we find 
 \ba
&\te q_{1} =\cc_1-2\cc_2 + 6 \cc_3 = 3072 \kk(  -\wa_{3,1}+\tfrac{1}{4}\,\wa_{3,2}) \ , \la{11.7} \\
 \te 
&\te q_{2} =\cc_1 - 4 \cc_2 = 3072\kk ( -  { 1 \ov 3}   \overline \wa_{2,1} +  \wa_{3,2}) \ . \la{1177}
\end{align}
We observe  that $q_1$ does not depend on $\wa_{2,i}$, {\em i.e.} 
 all $R^2$  corrections  obey $q_{1}=0$. They   should  thus preserve the 
$(1,0)$ supersymmetry of the boundary theory \ci{Kulaxizi:2009pz}. Indeed,  the $R^2$ terms  admit 
supersymmetric extension (such corrections appear, {\em e.g.}, in compactifications that break  half of maximal 
supersymmetry). The  $R^3$   corrections   consistent with $(1,0)$   supersymmetry should  obey 
\be \la{001}
(1,0): \qquad \qquad \wa_{3,2}= {4}\,\wa_{3,1}  \ . \ee 
Demanding   both $q_1$   and $q_2$  to vanish that   should correspond to the  maximal  supersymmetry case, 
{\em i.e.} to the $(2,0)$   6d  boundary theory,  gives\foot{The  terms  $\wS_{2,2},\wS_{2,3}, \wS_{3,5},\wS_{3,6},\wS_{3,7},\wS_{3,8}$  in \rf{11.3},\rf{11.4} 
 that on the leading-order  equations of motion ($R_{\m\n} = \Lambda g_{\m\n}$)   renormalize   only the cosmological constant term 
 should  also preserve the $(2,0)$  supersymmetry -- indeed,  their coefficients do not appear in $q_2$.
 The terms   that break  $(2,0)$  supersymmetry 
 even in the absence of  $\wS_{3,1},\wS_{3,2}$ and    are 
$\wS_{2,1}$    and  also $\wS_{3,3}$   and $\wS_{3,4}$  that reduce to $\wS_{2,1}$   on the leading-order  equations of motion.
}
\be 
(2,0): \qquad \wa_{3,2}= {4}\,\wa_{3,1}
\ , \ \ \ \ \ \  \qquad   \overline \wa_{2,1} = 12\wa_{3,1} \ .  \la{pp} \ee 

\iffa 
In this  case we get  
\ba \no 
& \cc_1=96 \cc \ , \ \qquad  \ \cc_2 = 24 \cc\ , \ \ \qquad  \cc_3 = - 8 \cc \ ,  \\
& \cc=- \kk (1  - 21  \overline \wa_{2,2} 
  - {222} \wa_{3,1} ) \ , \ \ \qquad \ \ \   \aa= -\kk( 1  - 21  \overline \wa_{2,2} 
    - {126} \wa_{3,1} )  \ . \la{xxx}
\end{align}
Assuming as   in \ci{Tseytlin:2000sf}  that we may apply  the purely-gravitational 
action \rf{11.2} 
to the  description   of 7d reduction of 11d   effective action  ({\em i.e.} assuming that possible flux-dependent terms  may be mimicked by 
``redundant" curvature   invariants)  we may apply these   expressions   to the 
$(2,0)$  theory where we  expect    to find   that (see  \rf{22},\rf{23}) 
\be \la{ri} \te 
\aa=  - {1\ov 72} N^3   (1 - { 9\ov 16 N^2} +...)    \ , \ \ \ \ \ \ \ \  \qquad  
\cc=  - {1\ov 72} N^3  (1 - {3 \ov 4  N^2} + ...)  \ .    \ee
These   values   follow  from \rf{xxx} if 
\be \la{aa1} \te 
\overline \wa_{2,2}  = { 27 \ov 1792 N^2} \ , \ \ \qquad\qquad    \ \ \ \wa_{3,1}=  { 1 \ov 512 N^2} \ .  \ee
In addition, we may  also  arrange   that  the  AdS  radius  is not renormalized   by 
fixing one of the   ``redundant" 
$R^2$  or $R^3$  coefficients in \rf{1117}   so that $ 3 f_2 = 2 f_3$. 
\fi 

\subsection{Special (1,0) case:   Lovelock  action} 

Let   us    study  in more detail    the $(1,0)$ case  \rf{001}.  This  constraint   implies   that the  two irreducible curvature 
invariants $I_{3,2}$ and $I_{3,1}$    appear in the same   combination as in the Euler density $E_6$. 
Indeed, introducing 
the higher order Euler densities
\foot{Here $\delta^{\nu_1... \nu_n}_{\mu_1 ...\mu_n} = n! \delta^{\nu_1} _{[\mu_1} ...  \delta^{\nu_n} _{\mu_n]} $
and we assume 
Euclidean signature, 
{\em i.e.} $\ep_d \,\ep_d\,  =   (d-2p)! \delta^{...}_{...} $.  
Note that $E_6$   used   in   \ci{Bastianelli:2000hi} 
 was   of the opposite sign.
 }
\be\la{1120}
E_{2p} \equiv  \delta^{\nu_{1}\nu_{2}\cdots\nu_{2p-1}\nu_{2p}}_{
\mu_{1}\mu_{2}\cdots\mu_{2p-1}\mu_{2p}}\,R^{\mu_{1}\mu_{2}}{}_{\nu_{1}\nu_{2}}\cdots
R^{\mu_{2p-1}\mu_{2p}}{}_{\nu_{2p-1}\nu_{2p}} =  \frac{1}{(d-2p)!}\,  \ep_d \,\ep_d\,  R\cdots R  \ , 
\ee
one finds  using  
 the explicit expressions  for $E_{2p}= \wa_{p,i} \wS_{p,i} $ in  the  bases  (\ref{11.3}) and (\ref{11.4})\foot{For summary of properties of $R^3$ invariants see, {\em e.g.}, \ci{vanNieuwenhuizen:1976vb,Metsaev:1986yb}.
 Note, in particular, that if  we introduce $J_1= \wS_{3,1}, \ J_2 = R_{\m\n\k\l} R^{\r\n\sigma\l} R_{\r\m\sigma\k} , \ J_3 = - \wS_{3,2}$ 
 then $J_2= J_3 +  { 1 \ov 4}  J_1$.
 Also, in the notation used in \ci{Bastianelli:2000hi}
 we have $A_{16}= J_1,\ A_{17}= J_2, \ A_{15}= - \wS_{3,3}, $ {\em etc.}
    } 
\iffa 
\be\la{1121}
 \te 
E_{4} : \quad  \wa_{2,i} = 4\cdot (1, -4, 1)\  , \qquad \qquad 
E_{6} : \quad  \wa_{3,i} =  16\cdot (1, 4, 12, {3\ov 2} , 12, 8, -6, {1\ov 2} )\ . 
\ee
\fi 
that  choosing $\Delta \LL$ in \rf{1122}  in the special  ``$E_4 +E_6$''   form   we  get 
\ba \te
&\te \Delta \LL_E ={ 1\ov 4}\,  L^2\,   \av_2\,  E_4  + { 1 \ov 16}\,  L^4 \, \av_3\,  E_6 \ , \ \ \ \ \ \ \ \la{rr1}  \\
&\te \wa_{2,i} = \av_2 \cdot (1, -4, 1) \ , \ \ \ \ \qquad    \wa_{3,i} =  \av_3 \cdot (1, 4, 12, {3\ov 2} , 12, 8, -6, {1\ov 2} ). \la{rr11}
\end{align}
Here   the coefficients in \rf{588} are 
\be \la{pp1} 
\overline \wa_{2,1} =\av_2 +   9 \av_3\ ,  \qquad   \ \  \overline \wa_{2,2} =3\av_2  - 15 \av_3\ ,  \qquad \ \ \  \wa_{3,2} = 4 \wa_{3,1} = 4 \av_3 \  .
\ee
For the particular case   of the action \rf{11.2} with $\Delta \LL$  in the Lovelock form  \rf{rr1}   treated as a complete  theory 
({\em i.e.} not  expanding   in $\av_2$ and $ \av_3$  and not including higher curvature terms) 
 the  corresponding  conformal  anomaly coefficients 
 were found in   \cite{deBoer:2009gx,deBoer:2011wk,Hung:2011xb}.   
  They can be expressed  as \ci{Hung:2011xb}
\def \qq  {{\rm f}} \def \rmq {{\rm f}} 
\ba
& \te \aa = -{\rm k} \, \qq^{-5/2} \,(1- 40 \av_{2}{\rmq}+ 180 \av_{3}{\rmq}^{2}), \quad\ \
\cc_{1} = - {96}\,{\rm k} \, \qq^{-5/2} \,(1     -  {104 \ov 3} \av_{2} \,{\rmq}\, +  68\,\av_{3}\,{\rmq}^{2}),  \no \\
&\te \cc_{2} =-24 \,{\rm k} \, \qq^{-5/2} \,(1-  {136 \ov 3}\av_{2}  \,{\rmq}\, +   100\,\av_{3} \,{\rmq}^{2}), \ \ \  \
 \cc_{3} = 8\,{\rm k} \,   \qq^{-5/2} \, (1-24\av_{2} \,{\rmq}\, +    36\av_{3} \,{\rmq}^{2}). \la{11221}
\end{align}
Here  $\qq$ is  a  function of $\va_2, \va_3$  given by a root of a cubic equation below. It    enters 
the expression for the renormalized AdS radius $r_* $ \foot{The action evaluated on AdS$_7$   is proportional to  
 $ r^7  [ - 42/r^2 + 30  +  70 \times 12  ( \va_2  r^2 -3 \va_3)/r^6 ] $
which  is  extremized on the   solution of     the above  equation for  $\qq \equiv 1/r^2 $ (we set  here  $L=1$).}
\be\la{1123}
r_*  =  {{\rmq}^{-1/2}}  \, {L}\ ,\qquad  \qquad {\rmq}-12\va_{2}\, {\rmq}^{2}+ 12\va_{3}\, {\rmq}^{3} =1\ ,
\ee
which generalizes  the  linearized expression in \rf{1118}. 
The expressions \rf{1122}  agree  with the leading order  results in \rf{588} after one uses \rf{rr1}, \rf{pp1} and expands to linear  order 
in $\av_2,\av_3$. 

From \rf{1122} we find that  the combinations of coefficients in \rf{11.7},\rf{1177} are 
\be \la{11233}
q_1 = 0 \ , \ \ \ \ \ \ \ \   \qquad \te  q_2 = -{1024} \,{\rm k} \,   \qq^{-5/2}  (\va_2\qq-  3 \va_3 \qq^2 ) \ ,  
\ee
{\em i.e.} $(1,0)$ supersymmetry is preserved but $(2,0)$  is broken   in general. 

Compared  to  the general case in \rf{1118},\rf{588} discussed  above 
where the $\aa,\cc_i$ coefficients depended   on 4 parameters and one more parameter  controlled  the 
deformation of the AdS radius, 
the Lovelock  Lagrangian  \rf{rr1}  depending  on just 2  parameters is too constrained. 
We may introduce an extra term (for example,   scalar $R^2$ or $R^3$)  
 to arrange that the AdS   radius is not renormalized, {\em i.e.}
$\qq=1$,  and then the coefficients in \rf{1122} will be linear  in $\av_2,\av_3$.

\iffa
At first order in   both $a_{2}$ and $a_{3}$  ({treating them on an equal footing}) 
we have ${\rm q} = 1+a_{2}+a_{3}+\dots$, and thus 
\be\la{1124}
\begin{split}
& \aa = -\frac{{\rm k}}{2}\,\Big(1-\frac{35}{6}\,a_{2}-\frac{35}{2}\,a_{3}+\dots\Big), \quad\ \ \ \ \ \
 \cc_{1} =-96\,{\rm k}\Big(1-\frac{97}{18}\,a_{2}-\frac{49}{6}\,a_{3}+\dots\Big), \\
&\cc_{2} = -24\,{\rm k}\,\Big(1-\frac{113}{18}\,a_{2}-\frac{65}{6}\,a_{3}+\dots\Big), \quad
 \cc_{3} = 8\,{\rm k}\,\Big(1-\frac{9}{2}\,a_{2}-\frac{11}{2}\,a_{3}+\dots\Big).
\end{split}
\ee
These are in  agreement  with    the general perturbative  expressions in  (\ref{11.5}) and (\ref{1118})
in the special case of  \rf{11211}. 
\fi

\iffa 
The $(2,0)$ case    $q_2=0$     is found when $\va_2$ and $\va_3$ are  correlated:   
 assuming   that   one can independently arrange to have  $\qq=1$   this  requires  $\va_2= 3 \va_3$, 
 {\em i.e.}  corresponds  to \rf{rr1}  of the form 
\be 
\te \Delta \LL ={ 1\ov 4}  L^2   \av_2 \big(   E_4  + { 1 \ov 12}  L^2    E_6 \big)  \ . \  \la{rr3}  \ee
This   combination by itself   does not reproduce the expected  value of $\aa$-coefficient 
  in  \rf{ri}   for $(2,0)$ theory, 
requiring an addition of at least one extra curvature invariant. 

The crucial point is that  irreducible  $RRR$ invariants  should   form the $E_6$ combination 
 in order to be consistent with supersymmetry  and thus   describe $(2,0)$   theory 
 was  originally  noticed  in  \cite{Tseytlin:2000sf}   where dimensional reduction of $R^4$     invariants in 11d 
  on $S^4$  was considered.\foot{The fact that $E_6$   contribution  preserves $(1,0)$  supersymmetry  was noted in \ci{deBoer:2009pn} 
but   further  connection to 11d and $(2,0)$  theory was not appreciated.} 
\fi

\def \e {\epsilon}

\subsection{$(2,0)$  case }

Let us now    go back to  the condition \rf{pp}   that should correspond to the (2,0)   theory. 
Assuming  \rf{pp} we  get  from \rf{588} (dots stand for possible higher order terms) 
\ba 
& \aa= -\kk( 1  - 21  \overline \wa_{2,2} 
    - {126} \wa_{3,1} ) + ... \ , \qquad \qquad  \cc=- \kk (1  - 21  \overline \wa_{2,2} 
  - {222} \wa_{3,1} ) + ...  \ ,  \la{xxxx}\\
  & \cc_1=96 \cc \ , \ \qquad  \ \cc_2 = 24 \cc\ , \ \ \qquad  \cc_3 = - 8 \cc \ . \la{yyy}
\end{align}
Assuming as  in \ci{Tseytlin:2000sf}  that we may use  the purely-gravitational  7d 
action \rf{11.2} 
for  the  description   of  the  $S^4$ 
 reduction of 11d   effective action  ({\em i.e.} assuming that possible flux-dependent terms  may be mimicked by 
``redundant" curvature   invariants)  we may apply these   expressions   to the 
$(2,0)$  theory where we  expect    to find   that (see  \rf{22},\rf{23}) 
\be \la{ri} \te 
\aa=  - {1\ov 288} \big[ 4 N^3   (1 - { 9\ov 16 N^2})  +  {7\ov 4} \big]  \ , \ \ \ \ \ \ \ \  \qquad  
\cc=  - {1\ov 288} \big[ 4 N^3  (1 - {3 \ov 4  N^2})  +  1  \big] \ .    \ee
As already mentioned in section 5,  here   the $N^3$   terms come for leading supergravity part in \rf{11.2}  \ci{Henningson:1998gx}, 
the constant terms (equivalent to the contribution of one tensor multiplet) 
come   from 1-loop 11d supergravity  contribution  \ci{Beccaria:2014qea}  while 
 the $1/N^2$ terms  in brackets should  
 represent the exact  contribution of  $R^4$   curvature corrections to the 11d M-theory 
action. 
These  terms   follow  from \rf{xxxx}  with $\kk$    given by \rf{117}  if 
\be \la{aa1} \te 
\overline \wa_{2,2}  = { 27 \ov 1792 N^2} \ , \ \ \qquad\qquad    \ \ \ \wa_{3,1}=  { 1 \ov 512 N^2} \ .  \ee
In addition, we may  also  arrange   that  the  AdS  radius  is not renormalized   by 
fixing one of the   ``redundant"  $u_{r,i}$ 
coefficients in \rf{1117}   so that $ 3 f_2 = 2 f_3$. 
That should    be important   to ensure that $\aa$ and $\cc$ do not receive   further  $1/N$   corrections
(depending on higher powers of $ \wa_{r,i}\sim { 1 \ov  N^2}$), cf. \rf{1123}.

In the case of  special Lovelock action \rf{rr1}  the (2,0) condition $q_2=0$ in \rf{11233} 
implies that  the coefficients  $\va_2$ and $\va_3$  should   be   related as  $\va_2= 3 \rmq \va_3$
and then  $\rmq$  solving cubic  equation in \rf{1123}   becomes  a non-polynomial function of $\va_3$. 
The same  will then apply to the conformal anomaly coefficients in \rf{1122}. 
To avoid this complication we   may assume that adding an extra   scalar  $R^2$ or $R^3$ invariant to \rf{rr1} 
one can independently arrange to have  $\qq=1$. In that case  we will have   $\va_2= 3 \va_3$,  and then 
  \rf{rr1}  will take  the form 
\be 
\te \Delta \LL_E ={ 1\ov 4}  L^2   \av_2 \big(   E_4  + { 1 \ov 12}  L^2    E_6 \big)  \ , \ \ \ \qquad  \ \ \ \ \ \ \ 
 \av_2 = 3 \wa_{3,1}=  { 3 \ov  512N^2} 
 \ ,  \la{rr3}  \ee
where we  fixed the value of $\av_2$ to match the value of $\cc$ in \rf{ri}   (see \rf{pp1},\rf{aa1}).
Note that this  correction      by itself   will  not reproduce the expected  value of $\aa$  in  \rf{ri}   
for any $\av_2$    but  the $\aa$-coefficient    will get a contribution from an extra $R^2$ or $R^3$ invariant  (cf. \rf{588}). 
 
The crucial point is that  irreducible  $R^3$ invariants in \rf{11.4}  should   form the $E_6$ combination 
 in order to be consistent with supersymmetry  and thus  potentially  describe  subleading correction
  to the conformal anomaly  of   (2,0)   theory   was  originally pointed out   in  \cite{Tseytlin:2000sf}.
There    it was suggested that    a  particular 
$R^4$ super-invariant  term in   11d action 
translates upon compactification on  $S^4$  into 
a  combination of $E_{6}$  and $E_{4}$ corrections to 7d action
 which should thus    produce    subleading   contributions  to the Weyl anomaly coefficients of (2,0) 
 theory.   While  the  discussion in \cite{Tseytlin:2000sf}    was admittedly heuristic
  (other possible 4-form and   Ricci  tensor terms    were  ignored      and  the contribution to
 the conformal  anomaly  coming  from the   $E_4$  term in 7d action was  not included)
 the key role of the $E_6$ invariant was  noticed. 
 
    To recall, the starting point in    \cite{Tseytlin:2000sf} 
was the    11d  $R^4$    invariant 
that  involves  the 8d  Euler density  $E_8 $  factor
(see   \rf{1120}).  
If $M^{11} = M^7 \times S^4$  then splitting indices in $4+7$ way and 
performing combinatorical count  one finds   that\foot{Here we correct  the coefficient  of the second term 
which was having an extra factor of 2   in  \cite{Tseytlin:2000sf}.
In this  second term 
we have to assign a 7d-type index to 
two out of four Riemann tensors. This may be done in $\binom{4}{2}=6$ ways, or $3\times 4$ divided by 
$2!$ because of the symmetry of the two $R$  factors.
}
\iffa  
 have 2 options  for each factor of the curvature which enter with   combinatoric coefficients 4  and $ 3 \times 4$: 
 $$
  \e_{11} \e_{11}  RRRR \to  4  \e_7 \e_7  R_7 R_7 R_7   \e_4 \e_4    R_4  + 12   \e_7 \e_7  R_7 R_7    \e_4 \e_4    R_4  R_4 $$
  Here  in the first term  we have one index of $\e_7$   uncotracted and   2 indices of $\e_4$ uncontracted while  
  in the second is vice versa.  Now   we need to  include $ { 1 \ov 3!} $    factor  to rearrange in terms of $E_{2n}$ 
  and note that  in the first term we will need only $1\ov 2!$  factor  so $1/3$ will remain, {\em i.e.}   we will get  naively 
  $E_8 (M^7 \times M^4) = {4 \ov 3} E_6(M^7) E_2(M^4) + 12   E_4(M^7)   E_4(M^4) $. 
  But what if we start   with  the form of 
  $E_{2n} = \delta^..._...  RRRR$?  Then by same argument we will   not have any  extra 3  involved. 
  Indeed, in the  case when we have chosen  $R_7 R_7 R_7   R_4 $ combination of curvatures 
  we still have $\e_{MNK a_1...a_6 m_1 m_2} \e^{MNK b_1...b_6 n_1 n_2 } $ where   $M,...$ are 11d  indices, $a,b$ are  7d ones, and $m,n$ are 4d ones.  then we need  to choose one  7d index among $MNK$ and that   gives  extra factor of 3. 
  Thus  
  \fi
  \def \ll {{\it  l}}
  \ba
 && E_8 (M^7 \times S^4)= {4 } E_6(M^7) E_2(S^4) + 6   E_4(M^7)   E_4(S^4) \no \\
 &&= { 3 \times 2^5\ov \ll^2}   E_6 (M^7)   +  { 3^2 \times 2^7\ov \ll^4}   E_4 (M^7)
\ , \la{02}  \end{align}
where $\ll$ is the radius   of $S^4$  which in the case   of AdS$_7 \times S^4$ solution\foot{
M5  brane metric is  $
ds^2= h^{1/3}(y)   ( dx_m dx_m)   +   h^{-2/3}(y)  dy_k dy_k $
where $
h= 1 + {q\ov y^3} , \   \ dy_k dy_k = dy^2 + y^2  dS_4$   
 so 
that  in the  near-horizon limit  we get 
$ds^2 = q(  y^{-1}  dx_m dx_m/y   + y^{-2} dy^2   + dS_4 ) $. 
The AdS part   is  $ 4  z^{-2} (dx_m dx_m +   dz^2)
$  were  $y={ 1 \ov 4}  z^2$   so  that the  AdS$_7$  radius is twice the radius of $S_4$. }
is 
$\ll = \ha L $   
where $L$ is the scale of AdS$_7$.
  Then  the resulting combination  in \rf{02} becomes 
\be \la{03} 
E_8 (M^7 \times S^4)  =   { 3^2 \times 2^{10} \ov L^4}   ( E_4   + \te  { 1 \ov 24}  L^2  E_6 )  \ . 
\ee
Note that  this combination of $E_4$ and $E_6$ is different from the  one  in \rf{rr3} 
that was apparently required to reproduce   the right (2,0)   ratio of the $\cc_i$-coefficients. 
This is not, however, a  contradiction as  some extra ``reducible"  (Ricci-tensor dependent) 
 $R^2$ or $R^3$ terms 
are required to be added to both \rf{rr3}   or \rf{03} 
   in order  to ensure that the  leading-order expression for the AdS$_7$ radius is not corrected  
   (which was an assumption in  \cite{Tseytlin:2000sf}). 
   As  we  have explained above,   starting with  a general enough combination of $R^2+R^3$ terms 
   one   can   reproduce  the values  of the conformal anomaly coefficients in \rf{ri}  and also ensure that 
   the AdS radius is not renormalized,   implying that the resulting coefficients do not get further $1/N^2$    corrections.

One    may attempt to  repeat similar  considerations 
 for   less supersymmetric  models  corresponding, {\it e.g.},  
 to  different  choices of $M^4$   compactification space or adding extra fluxes
 (see  \ci{Passias:2015gya} and refs. there). 
 In addition  to changing   the  coefficient 
 of the  $N^3$  in the leading terms as  the AdS  volume will be different   that  will also 
 lead to   subleading $R^2$    corrections that 
 may have coefficients  $N^2$   instead of $N$ as in \ci{HH}.

\iffa 
Let us  now compare it    with the one that appeared in \cite{Tseytlin:2000sf}.
 In that case
the correction to the 7d action is a combination of $E_{6}(\text{AdS}_{7})$ and $E_{4}(\text{AdS}_{7})$
up to Ricci and flux corrections needed to obtain a superinvariant. Neglecting these corrections, it is reasonable
that we are reproducing the above $(2,0)$ combination. 
There  the  starting point was the invariant $E_8(M^{11})$    on $M^{11} = M^7 \times M^4$ 
with special  case   being AdS$_7 \times S^4$ vacuum with radii 
$L_4 = {1 \ov 2} L$,  $L_7=L$. 
 Then  the claim was \foot{In \cite{Tseytlin:2000sf}   $L_4$ was denoted by $L$ and  $L_7$ was set to be equal to 1.} 
\be \la{444} 
E_8(M^{7}\times S^4) = {3^2 \cdot 2^7\ov  L^{4}_4} \Big(  E_4  +  { L_4^2    \ov 12}   E_6  \big)\ , 
\ee
where $E_4,E_6$   correspond to $M^7$.  
This thus disagrees   with  \rf{11.23}  by extra relative factor  $ 1 \ov 4 $.
$E_2 (S^4)  = { 1 \ov 2} \epsilon_{mnkl} \epsilon^{mnpq}  R^{kl}_{pq} = 4! /L^2_4$, 
$ E_4 (S^4)  =  \epsilon_{mnkl} \epsilon^{ab pq}  R^{kl}_{pq} R^{mn}_{ab}  = 4\times  4! /L^4_4$, 
so we end up with \rf{444}.  So relative coeff is 12 but  in units of $L_4$....
In \cite{Tseytlin:2000sf} the contribution of $E_4$ term to anomaly was (incorrectly) ignored
but as  we saw above it shifts $\aa$ and $\cc_i$ in general. 
\fi 
\iffa 
To see that this is indeed the case, we need to be careful about
normalizations. The 7d combination in \cite{Tseytlin:2000sf} is proportional to 
\be\la{1128}
X=e_{7}e_{7}\, RRR+ 2\, e_{7}e_{7}\,RR.
\ee
Using $e_d e_d  = (d-2p)! \delta\cdots\delta$ and comparing with the definition of $E_{2p}$ this is 
\be\la{1129}
X=(7-6)! E_{6}+2\,(7-4)! E_{4} = E_{6}+12\,E_{4},
\ee
proportional to  (\ref{11.23}) (here $L=1$).
\fi 
\iffa
{\bf old comments: } 
Regardless  this   disagreemeent, it is clear that  Ricci corrections are necessary because 
the above combination does not shift the a-anomaly whereas the main result of \cite{Tseytlin:2000sf} reads
\be\la{1130}
\mc A_{(2,0)} = -\frac{1}{\,3^{2}\,2^{5}}\Big[
(4N^{3}-\frac{9}{4}\,N)\,E_{6}+(4N^{3}-3N)\cdot 8\,(12I_{1}+3I_{2}-I_{3}) + \mc O(\nabla J)\Big].
\ee
We can consider the most general combination of quadratic and cubic terms (multiplied by a common coupling)
such that the type B correction has $Q_{1}=Q_{2}=0$, the AdS radius is not renormalized  and the relative corrections to the a-anomaly and c-anomaly
are $-\frac{9}{16\,N^{2}}$ and $-\frac{3}{4N^{2}}$ respectively. In addition, we want to correct the $E_{6}+12E_{4}$ structure by 
terms that vanish on Ricci flat background. The general solution is the following addition to the Einstein-Hilbert Lagrangian
\be\la{1131}
\frac{1}{N^{2}}\,\frac{1}{2^{13}}\Big[E_{6}+12\,E_{4}+\frac{8}{21}\,(\sum_{i=1}^{3}\wa_{2,i}\wS_{2,i}+\sum_{i=1}^{8}\wa_{3,i}\wS_{3,i})\Big],
\ee
with
\be\la{1132}
\begin{split}
(\wa_{2,i}) &=(0,\ k,\ -\frac{k+708}{7}) \\
(\wa_{3,i}) &= (0,\ 0,\ 7w, \ w, \ t_{1},\ t_{2},\ t_{3},\ -\frac{t_{1}+t_{2}+7t_{3}+214}{49})
\end{split}
\ee
with arbitrary $k,w,t_{1,2,3}$. Multiplying by $\frac{1}{2\kappa_{7}^{2}}=\frac{N^{3}}{3\pi^{3}}$ this gives
\be\la{1133}
S = -\frac{N^{3}}{3\pi^{3}}\int d^{7}x\sqrt{g}(R+30)-\frac{N}{3\cdot 2^{13}\cdot\pi^{3}}\int d^{7}x\sqrt{g}
(E_{6}+12E_{4}+\dots).
\ee
This must be compared with (4.5) of \cite{Tseytlin:2000sf} that is \footnote{
The prefactor of the subleading $\mc O(N)$ term has this origin. We refer to the numbering of equations 
of  \cite{Tseytlin:2000sf}. The coefficient of $E_{8}$ in $\mc I_{2}$ is, (3.2),(3.4), 
$b_{1}T_{2}(-2)(\frac{1}{4})$. Using $b_{1}=\frac{1}{(2\pi)^{4}3^{2}2^{13}}$ after (3.4), and $T_{2}=\frac{2N}{\pi}$ from (4.4), we get $-\frac{N}{2^{17}3^{2}\pi^{5}}$. This must be multiplied by the coefficient of 
$\epsilon_{7}\epsilon_{7}RRR$ in (4.3), where $L=\frac{1}{2}$ and also integrated over the sphere $S^{4}$ with 
$\text{vol}(S^{4}) = \frac{\pi^{2}}{6}$. Thus, the final coefficient of $E_{6}$ is
\be
\notag
-\frac{N}{2^{17}3^{2}\pi^{5}}\times \frac{3\cdot 2^{5}}{(1/2)^{2}}\times \frac{\pi^{2}}{6} = 
-\frac{N}{3^{2}\cdot 2^{11}\cdot \pi^{3}}.
\ee
}
\be\la{1134}
S = -\frac{N^{3}}{3\pi^{3}}\int d^{7}x\sqrt{g}(R+30)+\frac{\gamma\,N}{3^{2}\cdot 2^{11}\cdot\,\pi^{3}}\int d^{7}x\sqrt{g}
(E_{6})_{R_{MN}=0}.
\ee
There is agreement if $\gamma = -3/2^{2}$. 

\noindent
{\bf comments}: is the minus related to conventions in contractions ? What about the extra $1/2^{2}$. Does it come
from a mistake we did in the normalization of Euler densities ? Notice that the extra term in our case, like for instance 
$E_{4}$ do contribute to the anomaly. In general, as we know, terms that vanish on Ricci flat do nevertheless contribute
to the anomaly. So, are we expecting matching of $\gamma$ here with $\gamma=3$ in \cite{Tseytlin:2000sf}?

What I got from E8 is  the action on ads7

r^7 (30 - 42/r^2 + k  ( 12/r^4 -1/r^6 )   )

where    k is some  numerical coeff = 5160960

at the same time we have  in the file  the action

r^7  (30  - 42/r^2  +  70 ( a2/r^4  + 3 a3/r^6 )   )

so we need  a2 = - 36 a3   to match

that means  in general
  a2 E4/48 - 1/192 a3 E6
=  -  a3/192  (    144  E4  +   E6)

so  instead of 12 we have   extra 12 relative coeff.

In fact, this is actually  the combination that comes out  of eq 4.3
in my 00 paper where    i finally found a   mistake -- there should be
no  factor of 3 in the first term, {\em i.e.} it should read
 2^5/L^2 [   E6  + 36/L^2  E4 ]
so for L= 1/2  that    gives relative   144 coeff.
That confusing 3    absence means that gamma=3    claim   in my paper
goes back to most  natural gamma =1

that 3  came from  eq 4.1    that   has also a mistake --   -- this is
the reason for mistake in 4.3:
the first   factor should have 1/3   in  it, {\em i.e.} it should read
 E8 (M7 x M4) = 4/3    E2(M4)   E6 (M7)  +   12  E4(M4)  E4 (M7)

that has to do with   with coeff 1/3!    in E8 that has nowere to go
now as E7 (M7) has factor  1  and  E2(M^4) has factor  1/2   in front
of e_d e_d

I will  put explanations  into the file.
Not sure what   that tells us  -- this is not 2,0  combination.
But   I would still want to have  more checks on c_i values --
confusion in all these papers   suggests we should keep an open mind.
\fi 

\section*{Acknowledgments}
We  thank   M. G\"unaydin,  J. Heckman, C. Herzog, K. Intriligator,  
A. Parnachev,  A. Patrushev  and R. Roiban 
  for  useful discussions, comments  and questions. 
The  work of A.A.T. is supported by the ERC Advanced grant No.290456,
 the STFC  grant ST/J0003533/1
and also by  the  Russian Science Foundation grant 14-42-00047  associated with Lebedev Institute.

\appendix

\section{Conformal anomaly  of $V^{(1,0)}$  multiplet on Ricci flat background}
\label{A}

To provide  information  about $\cc_i$   coefficients  for higher derivative superconformal   vector multiplet  $V^{(1,0)}$,
here we compute the conformal anomaly \rf{4}  for  its  fields   in \rf{19} on Ricci
 flat background using the results of \cite{Bastianelli:2000hi}.
For a scalar, we have (using \rf{5} and dropping  total derivatives)
\be\la{a1}
7!\,   A_{6}(\varphi) =\te (\frac{5}{72}\,E_{6}-\frac{28}{3}\,I_{1}+\frac{5}{3}\,I_{2}+2\,I_{3})\big|_{R_{mn}=0}   \to \  \frac{28}{9}\,I_{1}+\frac{17}{9}\,I_{2} \ .
\ee
For the standard 6d  Majorana-Weyl  fermion 
\be\la{a2}
7!\,  A_{6}(\psi) =\te   (\frac{191}{288} E_{6}-\frac{224}{3}\,I_{1}-8\,I_{2}+10\,I_{3})\big|_{R_{mn}=0}   \to \ \frac{70}{9}\,I_{1}+\frac{29}{9}\,I_{2} \ .
\ee
From the explicit  form of the  4-derivative vector field  $V^{(4)}$   kinetic operator 
 on a curved background given 
in \ci{Beccaria:2015uta}   one can show that 
the  corresponding  partition function   on  a  Ricci flat background has the form 
\be\la{a3}
Z(V^{(4)}) = \Big[\frac{(\det \Delta_{0})^3}{(\det\Delta_{1})^2}\Big]^{1/2} \ , 
\ee
where $\Delta_{0,1}  = -\nabla^2$ are  Laplacians   defined on scalars and vectors 
respectively. 
The contribution of  $\Delta_{1}$  can be found in \cite{Bastianelli:2000hi}:
\be\la{a4}
7!\, A_6(\Delta_{1})\big|_{R_{mn}=0}  \te = -\frac{112}{3}\,I_{1}-\frac{50}{3}\,I_{2} \ .
\ee
Taking into account that  the  fermion in $V^{(1,0)}$   multiplet in \rf{19}  has
 $\slashed{\nabla}^{3}$ kinetic term  we   find for  the total contribution 
\be
\label{a5}
\begin{split}
 A_6(V^{(1,0)}) \big|_{R_{mn}=0} & = 
\Big(3\,A_6(\varphi)+2\cdot 3\cdot  A_6(\psi)+\big[2\,A_6(\Delta_{1})-3\, A_6(\varphi)\big]\Big) \big|_{R_{mn}=0} \\
&\te = -\frac{1}{360}\,(2\,I_{1}+I_{2}) = -\frac{1}{11520}\,  E_{6}\ .
\end{split}
\ee
Comparing to the  general  expression \rf{6}    this implies that the 
expected $(1,0)$   supersymmetry relation \rf{7} is satisfied. 
Furthermore,  using the  known \ci{Beccaria:2015uta}  value of $\aa$-coefficient 
 in Table \ref{T2} in we  conclude that $\cc_1 + 4 \cc_2 = { 62\ov 45}$, 
   in agreement with the values of $\cc_1,\cc_2$ in \rf{20}.

\section{Chiral  anomalies of  \csg   supermultiplet}
\label{C}

Here we shall  discuss   the computation  of  the chiral R-symmetry  anomaly  and 
the mixed  R-symmetry -- gravitational  anomaly 
coefficients $\a$ and $\b$  in \rf{11} for the \csg   multiplet  considered  in section 4, 
demonstrating the relation \rf{43}.

\subsection{General $p > 2$ case}
\label{C.1}

Let us  start with  coefficient $\a$   of   R-symmetry anomaly. 
We shall  consider the  anomaly  of $SU(2)$  subgroup of $USp(4)$ 
 R-symmetry    of fields  in Table \ref{T5}.
To fix normalizations,   for  a  positive-chirality  MW  fermion $\psi^+$ or ${1\ov 2}^+$  in a 
 representation $\mathbf d$  of $SU(2)_{R}$  we have (see, {\em e.g.}, \cite{Avramis:2006nb})
\be\la{c1}\te 
\mc I_{8}({1\ov 2}^+)  =  
-\frac{1}{48}\text{tr}\, F^{4} =- \frac{1}{3}\,Q^{(4)}_{\mathbf d}\,\big[c_{2} (SU(2)_R) \big]^{2} \ ,
\ee
where $Q^{(4)}_{\mathbf d}$ is the following sum of 4-th powers 
of $U(1)$ charges over the representation $\mathbf d$  (cf. \rf{33}) 
\be\la{c2}  
Q^{(4)}_{\mathbf d} = \sum_{k=0}^{d-1}\te 
\left(-\frac{d-1}{2}+k\right)^{4} = \frac{1}{240}\,d\,(d^{2}-1)\,(3\,
d^{2}-7) \ .
\ee
Thus the corresponding $\a$ coefficient in \rf{11}  is 
\be\la{c3}
\te  \alpha  ( { 1 \ov 2}^+) = -8\,Q_{\mathbf d}^{(4)} \ .
\ee
The  standard (gauge-invariant,  non-conformal)  MW gravitino 
contributes to the gauge anomaly  as 5 times  the  contribution of a MW    fermion  \ci{AlvarezGaume:1983ig}.  To find the  anomaly 
  of  a   conformal  massive (non-gauge) 
gravitino $\psi_m$  we need to add, as  in \rf{42},   the  contribution  of two  MW   fermions, {\em i.e.}
\be \la{c4}
\mc I_{8}(\psi^+_m) 
  \te =\mc I_{8}({ {3\ov 2}^+} )
  + 2 \,\mc I_{8}({{1\ov 2}^+} ) = 7 \,  \mc I_{8}({{1\ov 2}^+} ) \ , \ \ \ \ \ \qquad  
  \alpha  ( \psi^+_m) = -56\,Q_{\mathbf d}^{(4)} \ .
\ee
For a  selfdual  non-gauge 3-form  in  the same representation of $SU(2)_R$ 
 the required value  appears to be\foot{One  should  be able express the anomaly  of  the antisymmetric tensor
 in terms of a spinor field  anomaly as in  4d case \ci{Romer:1985yg}.}
 \be
\la{c5}  \te  \alpha (A^+_{mnk} ) = - 8\,  \alpha  ( { 1 \ov 2}^+) =       64\,Q_{\mathbf d}^{(4)} \ .
\ee
Given a representation $[a,b]$ of $USp(4)$ we can
decompose it into representations $(\mathbf{d_{1}}, \mathbf{d_{2}})$ of $SU(2)_{L}\times SU(2)_{R}$  and then  identify the right factor with the 
$SU(2)$  R-symmetry   subgroup   the anomalies of which we are to  compute. 
The  decomposition is   \cite{Couture:1980}
 \be\la{c6}
 [a,b] = \sum N_{d_{1},d_{2}}^{a,b}\,(\mathbf{d_{1}}, \mathbf{d_{2}}) \  ,
 \ee
 where the multiplicites $N_{d_{1},d_{2}}^{a,b}$ are   determined   by\footnote{
 For example, the  $USp(4)$  representations of 
 states in the  conformal supergravity multiplet ($p=2$  level in Table \ref{T5}) 
  decompose  as follows  
 \be
 \notag
 \begin{split}
& [0,2]  \to \bf (1,1)+(2,2)+(3,3)\ , \qquad
 [1,1]  \to \bf (1,2)+(2,1)+(2,3)+(3,2)\ ,\qquad  
 [0,1]  \to \bf (1,1)+(2,2)\ , \qquad \\
& [2,0]  \to \bf (1,3)+(2,2)+(3,1)\ , \qquad 
 [1,0]  \to \bf (1,2)+(2,1)\ , \qquad
 [0,0]  \to \bf (1,1).
 \end{split}
 \ee
}
 \be\la{c7}
 \sum N_{d_{1},d_{2}}^{a,b}\,
  n^{d_{1}-1}m^{d_{2}-1} = 
 \Big[
 (1-a\,n)(1-a\,m)(1-b)(1-b\,n\,m)
 \Big]^{-1}.
 \ee
 For a $USp(4)$ representation $[a,b]$   we get \ba
Q^{(4)}_{[a,b]} =&\sum N_{d_{1},d_{2}}^{a,b}\,d_{1}\,Q^{(4)}_{\mathbf d_{2}}\no  \\
=&\te  \frac{1}{6720}\,(a+1) (b+1) (a+b+2) (a+2 b+3) \big(2 a^4+8 a^3 b+16 a^3
+14
   a^2 b^2\no \\  &+52 a^2 b+29 a^2
    +12 a b^3+64 a b^2+74 a b-12 a+6 b^4+36
   b^3+44 b^2-30 b\big) \ . \la{c9}
\end{align}
Using the above relations,  the total contribution to $\a$ coefficient  from all the fields 
of \csg multiplet in Table \ref{T5}   is found to be 
\ba
\alpha(\text{CSG}_{p})
 = &-8\,(Q^{(4)}_{[1,p-1]}-Q^{(4)}_{[3,p-3]}+Q^{(4)}_{[3,p-4]}-Q^{(4)}_{[1,p-4]})\no  \\
  &-56\,(Q^{(4)}_{[1,p-2]}-Q^{(4)}_{[1,p-3]})  + 64\,(Q^{(4)}_{[0,p-1]}-Q^{(4)}_{[0,p-3]}) \no \\
\label{c10}
 = & -2\,\big[6p(p-1)+1\big]   \ ,
\end{align}
which   is   the   value quoted in  \rf{44}.

Similar analysis   can be repeated   for the  coefficient $\beta$ 
of the mixed  R-symmetry -- gravitational  anomaly in \rf{11}.  The analog of \rf{c1} here is 
\be \la{c11}
\te \mc I_{8}({1\ov 2}^+)  =  -\frac{1}{192} \text{tr}F^{2} \, \text{tr}R^{2}\,
= \frac{1}{24} \,Q^{(2)}_{\mathbf d}\,c_{2}(SU(2)_R)\,\rp_{1} \ ,
\ee
where $Q^{(2)}_{\mathbf d}$ is the  sum of squares  of   $U(1)$ charges 
 over an  $SU(2)_{R}$ representation 
of   dimension $d$
\be\la{c12}
Q^{(2)}_{\mathbf d} = \sum_{k=0}^{d-1}\te \left(-\frac{d-1}{2}+k\right)^{2} = \frac{1}{12}\,d\,(d^{2}-1)\ .
\ee
Thus  for a MW spinor in the representation $\bf d$ (cf. \rf{11})
\be
\la{c13}\te 
 \beta ( {1\ov 2}^+)  = - Q_{\mathbf d}^{(2)} \ .
\ee
The  standard  MW gravitino contributes to the 
mixed anomaly  with  a factor $-19$ compared to the MW spinor  \ci{AlvarezGaume:1983ig}. 
Again,  a massive conformal gravitino requires the addition of two extra  MW spinor
contributions,   so that the total $\b$  coefficient 
for a positive chirality  MW conformal massive gravitino  in representation $\bf d$ of $SU(2)_R$ 
is 
\be\la{c14} 
\beta (\psi^+_m)  = 17\,Q_{\mathbf d}^{(2)} \ .
\ee
The    self-dual non-gauge    antisymmetric   tensor  contribution  is
\be\la{c15} 
 \beta (A^{+}_{mnk})  =\te  16 \beta ( {1\ov 2}^+)  = -16 \,Q_{\mathbf d}^{(2)} \ .
\ee
Using that 
\be
\la{c16} 
\begin{split}
Q^{(2)}_{[a,b]} &=\te \frac{1}{240} (a+1) (b+1) (a+b+2) (a+2 b+3) \left(a^2+2 a b+4 a+2
   b^2+6 b\right) \ , 
\end{split}
\ee
the total contribution to $\beta$ from the  \csg  multiplet in Table \ref{T5}   is found to be 
\ba
\beta (\text{CSG}_{p}) =& -\,(Q^{(2)}_{[1,p-1]}-Q^{(2)}_{[3,p-3]}+Q^{(2)}_{[3,p-4]}-Q^{(2)}_{[1,p-4]})\no \\
&
+17\,(Q^{(2)}_{[1,p-2]}-Q^{(2)}_{[1,p-3]})
-16\,(Q^{(2)}_{[0,p-1]}-Q^{(2)}_{[0,p-3]}) \no \\
\label{c17}
 = & -\big[6p(p-1)+1\big] \ , 
\end{align}
in agreement  with \rf{44}. 

As a check on normalizations used above 
we can  re-derive the anomaly
polynomial of $\SS^{(1,0)}_{p}$ multiplet discussed in  section 3. 
According to Table \ref{T4}, 
the fields that contribute to chiral anomalies   there  are  
 a MW  spinor ${\ha}^+$  in the $SU(2)_R$ 
 representation  $\mathbf{p}$ and a MW  spinor  $\ha^-$ 
  in 
the representation $\mathbf{p-2}$, so that we find  
\be\no 
\te 
\SS^{(1,0)}_{p}: \quad \ \alpha = -8\,(Q^{(4)}_{\mathbf p}-Q^{(4)}_{\mathbf p-2}) = -(p-1)^{4}, \qquad 
\beta = -\,(Q^{(2)}_{\mathbf p}-Q^{(2)}_{\mathbf p-2}) = -\frac{1}{2}(p-1)^{2}, 
\ee
 in agreement with \rf{34}.

\subsection{(2,0)    conformal   supergravity}
\label{C.2}

The above discussion applied for  $p >2$. 
In the $p=2$  case  of  Table \ref{T5}  corresponding to the (2,0)  conformal supergravity multiplet the gravitino is massless    and 
thus requires a  special treatment. 
We recall the  conformal   representations of the  corresponding fields 
 in Table \ref{T6}. 
\begin{table}[H]
\be
\begin{array}{|l|l|}
\hline
 (\Delta; h_{1},h_{2},h_{3}) & USp(4) \\
\hline
(4;0,0,0) & [0,2] = \mathbf{14} \\
(\frac{9}{2};\frac{1}{2},\frac{1}{2},\frac{1}{2}) & [1,1] = \mathbf{16} \\
(5; 1,1,1) & [0,1] = \mathbf{5}\\
(5; 1,0,0) & [2,0] = \mathbf{10}\\
(\frac{11}{2};\frac{3}{2},\frac{1}{2},\frac{1}{2}) - 
(\frac{9}{2};\frac{1}{2},\frac{1}{2},\frac{1}{2})    & [1,0]= \mathbf{4} \\
(6;2,0,0) & [0,0] =\mathbf{1}\\
\hline
\end{array}
\nonumber
\ee
\caption{$SO(2,6)\times USp(4)$ representations of  fields of  (2,0)  conformal supergravity
} 
\label{T6}
\end{table}
Here in the gravitino entry we explicitly 
 indicated the subtraction of the   contribution of the 
 gauge  degree of freedom
 which  is the fermion of the same chirality. 
 
 The  gravitational anomaly count then gives\foot{Here  in the  conformal gravitino contribution we subtract the $\g$-trace  degree of freedom (which is the opposite  chirality fermion)
 from the   contribution of the standard  non-conformal gravitino ${3\ov 2}^+$, cf. \rf{42}.
 The  antisymmetric tensor is non-gauge one  so its  contribution  is  twice that of  the  gauge-invariant one  in \rf{40}.}
\be
\label{8.34}
\begin{split}
\te 
16\times \mc I_{8}({\ha}^+) +  5\times 2\,\mc I_{8}({T}^+) +
4\times \big[\mc I_{8}({3\ov 2}^+) + \mc I_{8}({\ha}^+) \big]= { 1 \ov 4!} \big( -\frac{13}{4}\,\rp_{1}^{2}+13\,\rp_{2}\big) \ . 
\end{split}
\ee
This agrees  with  the    general $p$   result \rf{86}  formally continued to $p=2$.

Similarly, we  may  compute the   chiral anomaly $\a$ and $\b$   coefficients
(the middle term coefficients here  are  the   conformal  gauge-invariant   gravitino  ones)
\be
\label{8.35}
\begin{split}
\alpha &= -8\,Q^{(4)}_{[1,1]}-{48}\,Q^{(4)}_{[1,0]}
+64\,Q^{(4)}_{[0,1]} = -26\ , \\
\beta &=  -Q^{(2)}_{[1,1]}+{18}\,Q^{(2)}_{[1,0]}-16\,Q^{(2)}_{[0,1]} = -13 \ . 
\end{split}
\ee
These are 
again in agreement with the  formal  $p=2$ continuation of  \rf{c10} and \rf{c17}.
 
\subsection{(1,0)    conformal   supergravity}
\label{C.3}

Let us now 
consider  the    chiral anomaly coefficients corresponding to (1,0)  conformal supergravity.
For comparison,  the  chiral fields   in the  multiplets of (2,0)  and (1,0)  conformal supergravities   are listed   
in Table~\ref{T7} below.
\begin{table}[H]
\be\no 
\def\arraystretch{1.3}
\begin{array}{|c|c|c|c|c|}
\hline
 \text{  } & G_R & \psi^{+} & A^{+}_{mnk} & \psi_{m}^{+} \\
 \hline
 (2,0) & USp(4) & [1,1]=\mathbf{16} & [0,1]=\mathbf{5} & [1,0]=\mathbf{4} \\
 \hline
 (1,0) & SU(2) &  \mathbf{2} & \mathbf{1} & \mathbf{2} \\
 \hline
\end{array}\notag
\ee
\caption{Chiral fields   in the  multiplets of (2,0)  and (1,0)  conformal supergravities. $G_{R}$ is the R-symmetry
group. $[a,b]$ is the representation of $USp(4)$ with Dynkin labels $a$, $b$.}
\label{T7}
\end{table}
It is then straightforward to find the   analogs of the (2,0)  CSG 
 expressions (\ref{8.34}) and (\ref{8.35}) in the (1,0)  CSG case: 
\be\la{c18}
\begin{split}
&\te 
2\times \mc I_{8}({\ha}^+) +   2\,\mc I_{8}({T}^+) +
2\times \big[\mc I_{8}({3\ov 2}^+) + \mc I_{8}({\ha}^+) \big]= { 1 \ov 4!} \big( -\frac{107}{80}\,\rp_{1}^{2}+{101\ov 20} \,\rp_{2}\big) \ , \\
& \te \alpha = -8\,Q^{(4)}_{\mathbf{2}}-{48}\,Q^{(4)}_{\mathbf{2}}
+64\,Q^{(4)}_{\mathbf{1}} = -7\ , \qquad 
\beta =  -Q^{(2)}_{\mathbf{2}}+{18}\,Q^{(2)}_{\mathbf{2}}-16\,Q^{(2)}_{\mathbf{1}} = \frac{17}{2}.
\end{split}
\ee
Thus the  set of 4 chiral anomaly coefficients here is 
\be\la{c19} \te 
\vec\alpha (\text{CSG}^{(1,0)}) = \big(-7, \frac{17}{2}, -\frac{107}{80},\frac{101}{20}\big) \ . 
\ee
Applying the relation \rf{12} of \ci{Cordova:2015fha}   to compute the $\aa$-coefficient 
we get 
\be \la{c20}
\aa(\text{CSG}^{(1,0)}) =\te   \frac{797}{3840} \ , 
\ee
which  agrees with  the value found in \ci{Beccaria:2015uta}   by an independent  method.
This provides a non-trivial  check of  consistency of  the  chiral anomaly values \rf{c19} 
and also of the relation \rf{12}.

\iffa 
A remarkable check of this result is the computation of the a-anomaly using Intriligator's formula. One obtains
\be
\aa[\text{CSG}^{(1,0)}] = \Big(\frac{16}{7},-\frac{16}{7},\frac{16}{7},\frac{6}{7}\Big)\cdot\Big(-7, \frac{17}{2}, -\frac{107}{80},\frac{101}{20}\Big)\,\aa(T^{(2,0)}) = \frac{
797}{3840},
\ee
\fi

Using  the relations \rf{14}   we may now  compute  the  corresponding $\cc_i$--coefficients:
\be 
\la{c21} \text{CSG}^{(1,0)}: \qquad 
\cc_1 =\te   \frac{157}{9} \ ,\qquad \ \ 
\cc_2 =\te   \frac{781}{180} \ ,\qquad \ \ 
\cc_3 =\te  -  \frac{263}{180} \ .\ee
One   may ask if  there  are  any  combinations 
of (1,0)  multiplets 
$\text{CSG}^{(1,0)}+k_{1}\, S^{(1,0)}+k_{2}\, T^{(1,0)}+k_{3}\, V^{(1,0)}$ 
that are  free of all chiral  (and thus  also  of conformal)  anomalies. 
The answer  turns out to be negative.  
The  chiral  gravitational anomalies  cancel ($\gamma=\delta=0$) for 
$k_{2}=10, \ \  k_{3}=k_{1}-13$.

\iffa   At this point we can look for combinations
\be
\vec\alpha[\text{CSG}^{(1,0)}]+k_{1}\,\vec\alpha[S^{(1,0)}]+k_{2}\,\vec\alpha[T^{(1,0)}]+k_{3}\,\vec\alpha[V^{(1,0)}]
\ee
with special properties. The above combination does not vanish for any choice of the  integer constants $k_{i}$.
Gravitation anomalies cancel ($\gamma=\delta=0$) for 
\be
k_{2}=10, \qquad k_{3}=k_{1}-13.
\ee
\ee 
If we also impose that the a-anomaly is zero, there is only the unphysical solution 
\be
k_{1}=-8, \qquad k_{2}=10, \qquad k_{3} = -21.
\ee
\fi

\section{Casimir energy of  6d  supermultiplets}
\label{D}

 One  may wonder if the expression  for  the    Casimir energy  $E_c$ on $S^5$    for (1,0) 
 superconformal  6d theories can also   be expressed, like the conformal anomaly coefficients $(\aa, \cc_i)$
 in \rf{13}, \rf{14} 
  in terms of the   chiral  anomaly coefficients 
 $\vec\alpha =(\a,\b,\g,\d)$ in \rf{11}.\foot{Here we are consider
   the standard Casimir energy, not the ``supersymmetric" one  in  \ci{Assel:2015nca,Bobev:2015kza}  and refs. there.}

   Indeed,  $E_c$  is determined by the $T_{00}$    component of stress 
 tensor and should thus   be related \ci{Cappelli:1988vw,Herzog:2013ed,Huang:2013lhw}
 to  the   $\aa$-coefficient  and  also
   to a combination of  total derivative  term coefficients 
  in conformal anomaly.  The latter may be expected to be  more constrained 
  in the  supersymmetric case  and   may again  be  related to the  chiral anomaly coefficients. 
  One   should  note, however, that,    in  general,
   $E_c$  (and  the derivative terms in the  trace  anomaly)  
  are  scheme-dependent 
  so  that a relation  to scheme-independent  chiral anomaly coefficients    may hold only in a  particular 
  ``supersymmetric" 
  scheme.

  The   Casimir energies   for individual  6d conformal   fields can be computed  as in 
  \ci{Beccaria:2014qea}.
  Assuming  that $E_c$  (being related to the trace of  stress tensor)     may be  given by  a linear combination 
  of  the chiral anomaly coefficients  and   using 
   particular examples of (1,0)   supermultiplets   as data points  we have  found the  following  
 expression  for $E_c$ in terms of the  coefficients in  in the anomaly polynomial \rf{11}
  \be \la{c30}
  E_c = \te -\frac{1}{4}\big( \a-  \frac{31}{24} \b   +\frac{5}{3}\g   + \frac{29}{48} \d\big)  \ . 
  \ee 
  Note that in contrast to the conformal anomaly coefficients in \rf{12},\rf{14} 
 $E_c$  depends   not only on  $\a-\b$  but also   on  $\a+\b$.  
 This  relation is in agreement with  the values for $E_c$  for particular (1,0) 
 multiplets in \rf{18}, \rf{19}, \rf{31}  which  can be   found directly 
 \ba
\label{c31} \te 
&E_{c}(S^{(1,0)}) \te = -\frac{37}{3840}, \qquad 
E_{c}(T^{(1,0)}) = -\frac{71}{1280}, \qquad
E_{c}(\text{CSG}^{(1,0)}) = \frac{16471}{3840} \ , \\
&E_{c}(\SS^{(1,0)}_{p}) \te = \frac{1}{4}(p-1)^{4}-\frac{31}{192}(p-1)^{2}+\frac{37}{3840} \  , \la{c32}\\
& \la{c33}
E_{c}(\text{CSG}_{p})\te  = -2 \big[ 6 p (p-1) +1 \big] E_{c}(T^{(2,0)}) 
\ , \qquad \qquad   E_{c}(T^{(2,0)}) = - { 25\ov 384} \ . 
\end{align}
 In the (2,0)   case where    we  expect to   have $\b= 4 \g =- \d$  (see  \rf{16},\rf{166}) 
  so that $\aa$-coefficient  is given  by \rf{17} we get  for $E_c$ in  \rf{c30} 
 \be 
 E_c = \te -\frac{1}{4}\big( \a +  \frac{71}{48} \d \big)   \ . \la{c323}
 \ee 
 In contrast   to what happens in  the maximally supersymmetric case in 4  dimensions 
  this  $E_c$  is not directly proportional to  the corresponding expression 
   for $\aa= - {1\ov 72} ( \a  + { 9 \ov 8} \d)$ in \rf{17}.
   This   suggests   that  in 6d  case   the derivative term 
   contribution to the relation between $E_c$ and $\aa$-anomaly 
    does not vanish even in the (2,0)  case (cf. \ci{Herzog:2013ed}).
   

\iffa 
We do not include $\text{CSG}^{(1,0)}$ because of the  proportionality of its anomaly polynomial 
with that of $T^{(2,0)}$. In addition, we can compute with our old 6d expressions the Casimir energy 
of $S^{(1,0)_{p}}$ with the result
\be
\label{10.2}
E_{c}(\SS^{(1,0)}_{p}) = \frac{1}{4}(p-1)^{4}-\frac{31}{192}(p-1)^{2}+\frac{37}{3840}.
\ee
We can look for four coefficients $e_{i}$ such that 
\be
\label{10.3}
E_{c} = (e_{1}, e_{2}, e_{3}, e_{4})\cdot \vec\alpha.
\ee
This is non trivial. Indeed, using only the data in (\ref{10.1}), we obtain the following expression depending
on the single parameter $e_{1}$
\be
\vec e = \
\Big(e_1,\ \frac{77}{288}-\frac{2}{9}\,e_{1},\ \frac{32}{27}\,e_{1}-\frac{13}{108},
\ \frac{56}{27}\,e_{1}+\frac{635}{1728}\Big).
\ee
Then, we obtain for $\SS^{(1,0)}_{p}$, 
\be
\begin{split}
E_{c}(\SS^{(1,0)}_{p}) &= -e_{1}\, (p-1)^4+ \left(
\frac{1}{9}\,e_{1}-\frac{77}{576}\right) \,(p-1)^{2}+\frac{37}{3840}
   \end{split}
   \ee
 Comparing with (\ref{10.2}), we see that there is agreement of all terms for $e_{1}=-\frac{1}{4}$.
 Hence, we have found
 \be
 \vec{e} = \Big(-\frac{1}{4}, \ \frac{31}{96}, \ -\frac{5}{12}, \ -\frac{29}{192}\Big).
 \ee
 With this choice, the Ansatz (\ref{10.3}) reproduces the correct Casimir energy for all known free multiplets.
 \fi


\bibliography{HigherSpin6d-Biblio}
\bibliographystyle{JHEP}

\end{document}